\IfFileExists{revtex4-2.cls}{
\newcommand*{\revtexcls}{revtex4-2}
}{
\newcommand*{\revtexcls}{revtex4-1}
}
\newcommand*{\ARXIV}{}
\ifdefined\ARXIV
\documentclass[twocolumn,amsmath,amssymb,showpacs,aps,longbibliography]{\revtexcls}
\else
\documentclass[twocolumn,amsmath,amssymb,showpacs,aps,longbibliography,prl]{\revtexcls}
\fi
\usepackage[bookmarks=false,linkcolor=blue,urlcolor=blue,colorlinks,citecolor=blue]{hyperref}
\usepackage{graphicx}
\usepackage{bm}
\usepackage{amssymb}
\usepackage{amsmath}
\usepackage{amsfonts}
\usepackage{dsfont}
\usepackage{mathtools}
\usepackage{nicefrac}
\usepackage{paralist}
\usepackage[rgb]{xcolor}
\usepackage{mathrsfs}
\usepackage{tabularx}
\usepackage{xspace}
\ifdefined\ARXIV
\else
    \usepackage{xr}
\fi

\newcommand{\eg}{\emph{e.g.}\xspace}

\newcommand{\etc}{\emph{etc.}\xspace}

\def\multiset#1#2{\ensuremath{\left(\kern-.3em\left(\genfrac{}{}{0pt}{}{#1}{#2}\right)\kern-.3em\right)}}

\newtheorem{corollary}{Corollary}
\newtheorem{theorem}{Theorem}

\newtheorem{claim}{Claim}

\newcommand{\w}{{\mathbf w}}
\newcommand{\s}{{\mathbf s}}

\newcommand{\A}{{\mathcal A}}

\newcommand{\X}{{\mathcal X}}

\newcommand{\C}{{\mathbb C}}

\newcommand{\R}{{\mathbb R}}
\newcommand{\N}{{\mathbb N}}

\newcommand{\abs}[1]{\left\lvert#1 \right\rvert}
\newcommand{\norm}[1]{\left\|#1 \right\|}
\newcommand{\indc}[1]{\mathds{1}\left[#1\right]}

\newcommand{\shortminus}{\scalebox{1.5}[1]{-}}

\DeclareMathOperator{\softplus}{softplus}
\DeclareMathOperator{\sgn}{sgn}

\begin{document}

\title{Neural tensor contractions and the expressive power of deep neural quantum states}

\author{Or Sharir}
\email{or.sharir@cs.huji.ac.il}
\affiliation{The Hebrew University of Jerusalem, Jerusalem, 9190401, Israel}
\author{Amnon Shashua}
\email{shashua@cs.huji.ac.il}
\affiliation{The Hebrew University of Jerusalem, Jerusalem, 9190401, Israel}
\author{Giuseppe Carleo}
\email{giuseppe.carleo@epfl.ch}
\affiliation{Institute of Physics, École Polytechnique Fédérale de Lausanne (EPFL), CH-1015 Lausanne, Switzerland}

\begin{abstract}
We establish a direct connection between general tensor networks and deep feed-forward artificial neural networks. The core of our results is the construction of neural-network layers that efficiently perform tensor contractions, and that use commonly adopted non-linear activation functions. The resulting deep networks feature a number of edges that closely matches the contraction complexity of the tensor networks to be approximated. In the context of many-body quantum states, this result establishes that neural-network states have strictly the same or higher expressive power than practically usable variational tensor networks. As an example, we show that all matrix product states can be efficiently written as neural-network states with a number of edges polynomial in the bond dimension and depth logarithmic in the system size. The opposite instead does not hold true, and our results imply that there exist quantum states that are not efficiently expressible in terms of matrix product states or PEPS, but that are instead efficiently expressible with neural network states.  
\end{abstract}

\maketitle

\paragraph*{Introduction --}
Many fundamental problems in science can be formulated in terms of finding an explicit representation of complex high-dimensional functions, ranging from time-dependent vector fields to normalized probability densities. 
In recent years, Machine Learning (ML) techniques based on deep learning~\cite{goodfellow_deep_2016} have become the leading numerical approach for approximating high-dimensional functions found in industrial applications.
Due to this success, ML methods have also been recognized as a prime computational tool to attack functional approximation problems in physics~\cite{carleo_machine_2019}. 

In quantum physics, one of the main theoretical challenges in describing interacting, many-body systems stems from the complexity of finding explicit representations of many-particle quantum wave functions.
Tensor networks states~(TNS) are a well-established general-purpose ansatz for representing such functions. 
TNS are intrinsically rooted in the notion of locality in quantum systems and constitute both a key theoretical language to analyze many-body phenomena as well as a powerful numerical tool for simulations \cite{white1992density,schollwock2011density,orus_tensor_2019,verstraete2008matrix,cirac_matrix_2020}.
Recently, neural-network-based representation of quantum states, dubbed NQS, have been introduced~\cite{carleo2017solving} and subsequently used in a variety of variational applications.
A key theoretical question is how these two alternatives  relate to each other, and whether some families of quantum states are better described in terms of one of them. 

Several theoretical properties of NQS have been established to date. General representation theorems for neural networks \cite{Cybenko:1989fm} guarantee that sufficiently large NQS can describe arbitrary quantum states. Moreover, exact representations of many-body ground states of local Hamiltonians can be analytically found in terms of deep Boltzmann Machines \cite{carleo_constructing_2018}. Both representation results however do not bound the size of the corresponding NQS networks that, in the worst case, can be exponentially large in the number of physical degrees of freedom \cite{gao2017efficient}. 
Despite the worst-case exponential bound on NQS, examples of physically-relevant quantum states that can be efficiently represented are numerous. These encompass both analytical and numerical results. On the analytical side, for example exact and compact NQS representations of several correlated topological phases of matter are known \cite{deng_machine_2017,kaubruegger_chiral_2018,glasser_neural-network_2018,lu_efficient_2019}. On the numerical side, suitable learning algorithms have shown competitive results to find ab-initio approximate description of many physical systems of interest in physics \cite{PhysRevX.8.011006,choo_two-dimensional_2019,sharir_deep_2020,schmitt_quantum_2020,hibat-allah_recurrent_2020,torlai_neural-network_2018} and chemistry \cite{pfau_ab_2020,hermann_deep-neural-network_2020,choo_fermionic_2020}. 

As mentioned, a well-established paradigm for describing many-body quantum states are TNS. While generic TNS are widely believed to be general enough to compactly describe most physical quantum states, however only a restricted subset of them are amenable for numerical calculations. A determining factor in the applicability of TNS as variational quantum states is played by how complex it is to use these representations to compute physical quantities, and it is in turn related to the complexity of contracting TNS.  TNS that can be efficiently contracted most notably encompass matrix product states (MPS) \cite{white1992density}, a very powerful representation of low-entangled states in one-dimension. Higher-dimensional TNS are in general to be contracted only approximately, and rigorous complexity results have been established. For example, computing expectation values of physical quantities over planar tensor networks in two dimension, the Projected Entangled Pair States (PEPS) \cite{verstraete2004renormalization}, is non polynomial problem that is known to belong to the \#P complexity class \cite{schuch_computational_2007,haferkamp_contracting_2020}. 

Given the distinctive features of NQS and TNS, several works have studied possible connections between the two representations.
For example, the volume-law entanglement capacity of neural networks has been established in several works  \cite{deng2017quantum,PhysRevB.97.085104,levine2019quantum}. 
Also, mappings between the two classes of states have been realized, including between general fully-connected NQS and MPS with exponentially large bond-dimension \cite{PhysRevB.97.085104}. An approach mapping MPS onto non-standard neural-networks has also been introduced \cite{pastori_generalized_2019}. Despite the important theoretical progress, however a direct mapping between generic, efficiently contractible TNS and standard NQS has not been established to date. This situation for example leaves open the possibility that TNS can offer a general representational advantage over NQS representations \cite{borin_approximating_2020,park_are_2020}, and that there might exist compact, contractible TNS that cannot be expressed by means of compact NQS.   

In this work, we establish a direct mapping between TNS in arbitrary dimension and NQS. By directly constructing neural-network layers that perform tensor contractions, we show that efficiently contractible TNS can be constructed in terms of polynomially sized neural-networks. Our result, in conjunction with previously established results on the entanglement capacity of NQS, then demonstrates that NQS constitute a very flexible classical representation of quantum states, and that TNS commonly used in variational applications are strictly a subset of NQS.  

\begin{figure}
    \centering
    \includegraphics[width=\linewidth,keepaspectratio]{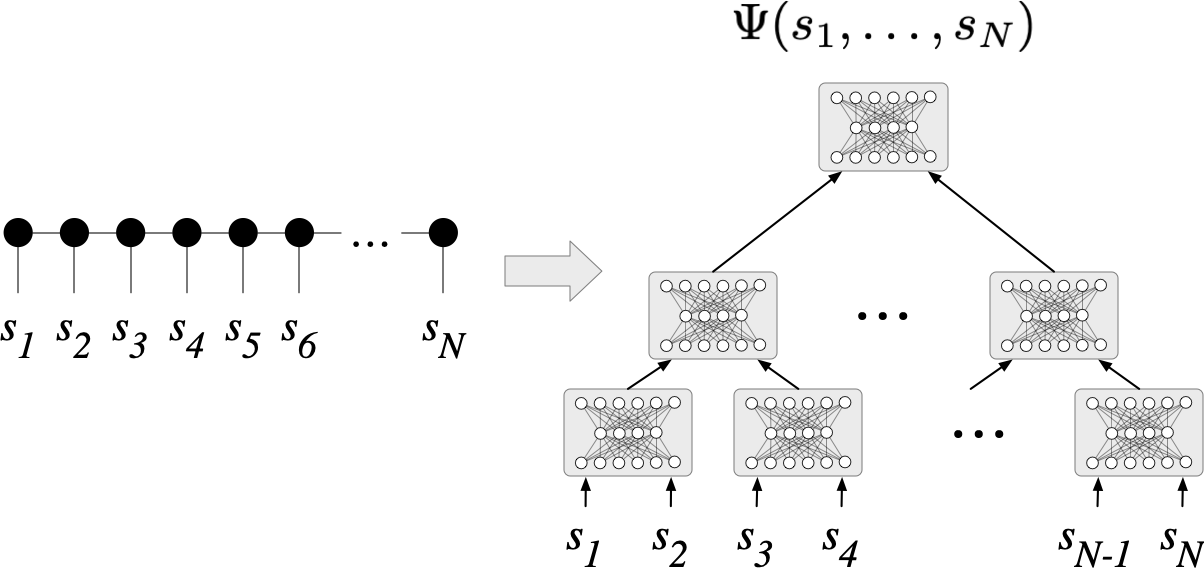}
    \caption{\label{fig:mps_to_nqs}We demonstrate a mapping from any tensor network with an efficient contraction algorithm to a compact neural network. In this figure we illustrate our coarse-grained construction of a Neural Network $\epsilon$-approximation of a Matrix Product State over $N$ sites, each of $d$ degrees of freedom, and bond dimension $\chi$. The resulting neural network is of depth $\tilde{O}\left(\ln N {+} \sqrt{\nicefrac{1}{\epsilon}}\right)$ and uses only $\tilde{O}\left(N (d {+} \chi)\chi^2 {+} \sqrt{\nicefrac{1}{\epsilon}}\right)$ edges.}
\end{figure}

\paragraph*{Preliminaries --}
We consider in the following a pure quantum system, constituted by $N$ discrete degrees of freedom ${\s{\equiv}(s_1,\ldots,s_N)}$ (\eg spins, occupation numbers, \etc) such that the wave-function~(WF) amplitudes $\langle s | \Psi\rangle = \Psi(\s)$ fully specify its state. Following the approach introduced in \cite{carleo2017solving}, we can represent $\log(\Psi(\s))$ as $g_1(\s) + i \cdot g_2(\s)$, where $g_1$ and $g_2$ are two outputs of a feed-forward neural network, parametrized by a possibly large number of network connections.
Given an arbitrary set of quantum numbers, $\s$, the output value computation of the corresponding NQS can generally\footnote{This is the classical definition of a neural network. However, some of the models used today slightly deviate from it, e.g., self-attention modules use bilinear operations in addition to affine ones. While our proofs for the main results only consider the classical definition, extending it to support many of these variants is trivial.} be described as two roots of a directed acyclic graph $(V, E)$, where the value of each node $v \in V$ is recursively defined by $v(\s) = \sigma\left(b_v + \sum_{(u, v) \in E} W_{u, v} u(\s) \right)$, where $\left\{W_{u, v} \in \R\right\}_{(u,v) \in E}$ and $\left\{b_v \in \R\right\}_{v \in V}$ are the parameters of the network, and $\sigma: \R \to \R$ is some non-linear function known as the \emph{activation function}, e.g., $\mathrm{ReLU}(x) = \max(x, 0)$ or $\softplus(x) = \log(\exp(x) + 1)$ \citep{ReLU,Softplus}. The root nodes of the network can optionally use the identity instead of a non-linear activation function. The depth of a neural network is defined as the maximal distance between an input node and the roots.

Alternatively, a state $\Psi(\s)$ can also be viewed as a complex tensor $\A_{s_1, \ldots, s_N}$ that is in turn represented in terms of tensor factorization schemes. Most forms of tensor factorizations are conveniently described graphically via Tensor Networks~(TN), undirected graphs whose nodes are tensors and edges specify contractions between connected tensors. See App.~\ref{app:tn_intro} for a brief introduction to TN.
In the next section we will present our main results on efficiency of approximating TN by NN.
To properly discuss the complexity of computing a TN, we have to be specific on how a given TN is computed.
First, a contraction order must be selected, i.e., the order by which intermediate tensors are computed (see App.~\ref{app:tn_intro} for a precise description).
Second, we must precisely describe the computational circuit of a given TN to be able to characterize some structural properties, e.g. depth and number of neurons, of the NN approximating it.
Given a contraction order, the value of $\Psi(\s)$ can alternatively be described in the form of an \emph{arithmetic circuit}, i.e., a computational graph comprising product and weighted-sum nodes.
Specifically, the value for a product node $v \in P$ is given by $v(\s) = \prod_{(u, v) \in E} u(\s)$, and for a weighted-sum node $v \in S$ is given by $v(\s) = \sum_{(u,v) \in E} W_{u, v} \cdot u(\s)$, where $\left\{W_{u, v} \in \C\right\}_{(u,v) \in E}$ are the parameters of the circuit, corresponding to the tensor nodes in the tensor network.
Input to the arithmetic circuit is represented by leaf input nodes, where for every $s_i$ and possible value $k$ there is an indicator node $v_{i,k} = \indc{s_i = k}$.
The depth of the circuit is defined the same as for neural networks.
See Fig.~\ref{fig:mps_contraction} for an illustration of a simple TN to AC conversion.

\begin{figure}
    \centering
    \includegraphics[width=0.5\linewidth,keepaspectratio]{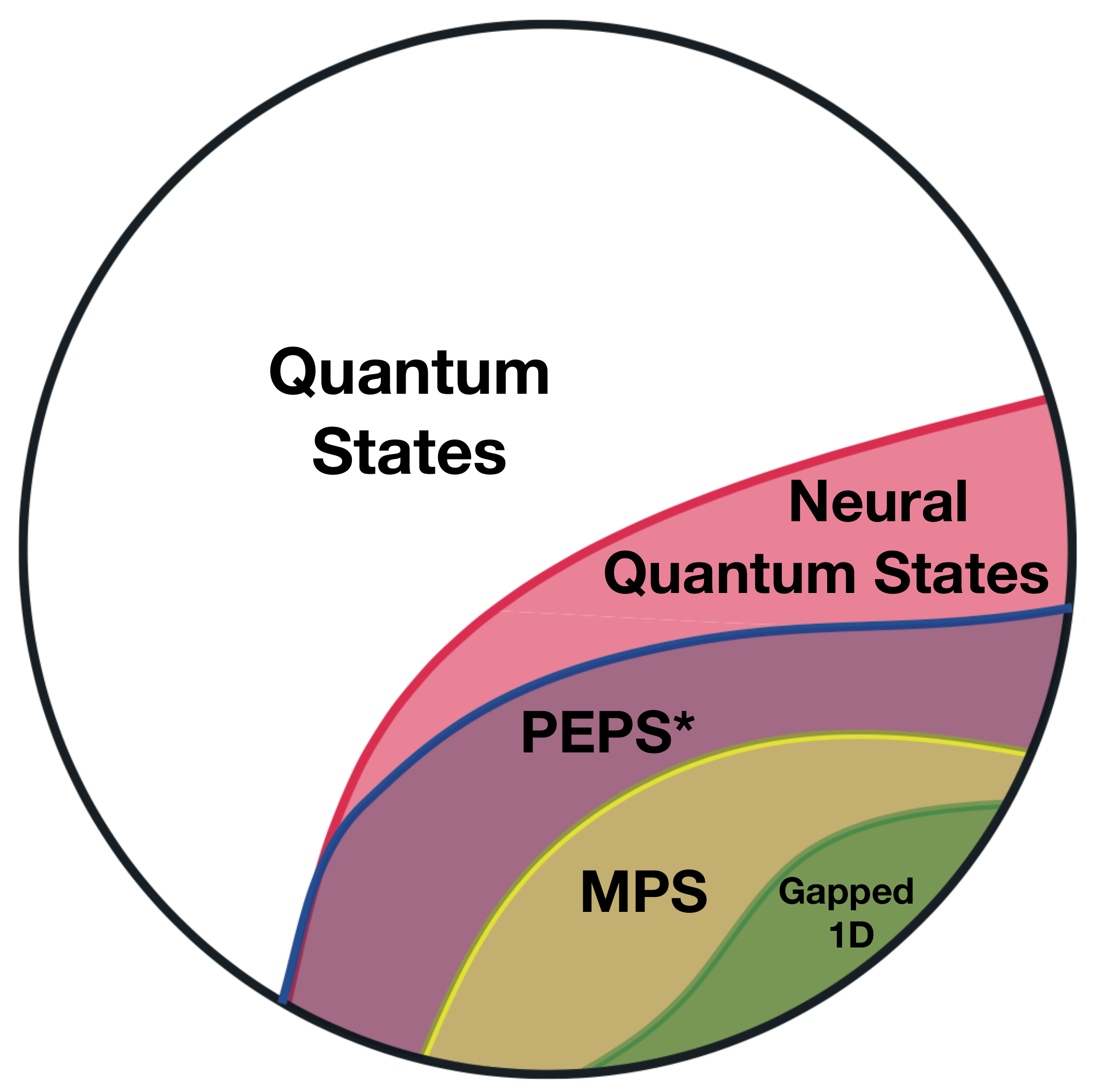}
    \caption{\label{fig:diagram} Expressive power of classically tractable variational quantum states. Different classes of quantum states describing a qudit system with $N$ degrees of freedom and comprising $\mathrm{poly}(N)$ variational parameters are compared. Matrix Product States (MPS) can efficiently represent gapped ground-states of one-dimensional systems. PEPS* denotes here Projected Entangled Pair States of bond dimension $\chi$ that are exactly or approximately contracted in $\mathrm{poly}(N,\chi)$ time on a classical computer. Neural Quantum States (NQS) comprise all polynomially tractable TN, thus include MPS, and $\mathrm{PEPS}*$, while also representing additional states with volume law entanglement that are not efficiently described by such planar TN.}
\end{figure}

\paragraph*{Main Results --}
Here we present our main results. First, that NN can represent any quantum state that is modeled by a TN with the same efficiency. Second, that there exist states that NN can model efficiently, but require exponential time for common forms of TN. The main outcome of our work is the representability diagram in Fig. \ref{fig:diagram}, summarizing the expressive power of NN and TN as variational quantum states.  
As discussed, the expressive efficiency of TN is defined with respect to a given contraction scheme that gives rise to an explicit computation in the form of an AC, composed of product and weighted sum operations. Hence, the fundamental question is whether AC can be efficiently simulated by NN. 

While the exact relationship between NN and AC has not been well studied~\citep{cohen2018analysis}, several works did study the relationship between NN and other polynomial functions~\citep{Poggio2017,Yarotsky2017,telgarsky17a}.
However, the prior methods do not yield sufficiently good bounds when applied to the problem at hand, resulting in impractical results. This inefficiency is inherently related to focusing on linear metrics between functions, rather than multiplicative. Because WF amplitudes are normalized, their absolute values are very small while their relative values are often orders-of-magnitude apart. See App.~\ref{app:related_theoretical_results} for a longer discussion.

As opposed to prior approaches, we consider the approximation of the log-value of AC, i.e., finding $g$ such that $\norm{g-\ln f}_\infty < \epsilon$~--~which translate to multiplicative bound in linear space~--~rather then $\norm{g-f}_\infty < \epsilon$.
Working in log-space has the advantage that more reasonable values (not dependent on $N$) for $\epsilon$ are sufficient for a meaningful approximation of WF amplitudes. We use the infinity norm to measure the error of two states because it gives a precise estimate over all inputs. Another common measure for the closeness of two quantum states is their fidelity, i.e., $F(\psi, \phi) = \frac{\abs{\langle \psi | \phi \rangle}^2}{\langle \psi |\psi \rangle \langle \phi | \phi \rangle}$. However, notice that closeness of the log-value under the infinity norm also implies closeness under the fidelity, while infinity norm in the linear domain does not entail such relationship unless $\epsilon$ is very small (see App.~\ref{app:fidelity} for proofs):
\begin{claim}\label{claim:good_fidelity}
   Let $\psi$ and $\phi$ be wave functions such that $\norm{\ln \psi -\ln \phi}_\infty < \epsilon$, then $F(\psi, \phi) \geq 1-\epsilon^2$.	
\end{claim}
\begin{claim}\label{claim:bad_fidelity}
	There exist wave functions $\psi,\phi:2^N \to \C$ such that $\norm{\psi - \phi}_\infty < \epsilon$, but $F(\psi, \phi) < 2^{-N}$ when $\epsilon^{-2} \ll 2^N$.
\end{claim}

We assume the magnitude of the AC's output is strictly positive for all inputs and greater than some fix value, $f_{\min}$, such that the log-value is well-define.
$f_{\min}$ can be extremely small, on the order of ${10}^{-{10}^{10}}$, without having a meaningful impact on our results and so it bares little effect in practice.
Furthermore, to simplify the presentation of our proofs, we assume the absolute value of both real and imaginary parts to be strictly positive, though this last assumption could be relaxed.
Under this settings, we proved that NN can simulate AC to almost arbitrary precision and with little overhead:

\begin{theorem}\label{th:main}
Let $f:\X \to \C$ be a complex-valued function given by an arithmetic circuit comprising $n$ nodes and $m$ edges, of depth $l$, and using complex parameters. Assume $0 < f_{\min} \equiv \inf_{x \in \X} \min\{\abs{\mathrm{re}(f(x))},\abs{\mathrm{im}(f(x))}\}$, and define $W_{\max} \equiv \max\{1, \max_{e \in E} W_e\}$. Then, there exist a function $g:\X \to \R^2$ described by a neural network comprising $O\left(n + m + c\right)$ nodes, $O\left(m + c\right)$ edges, of depth $O\left(l \log(m) + c\right)$, and using softplus activation functions and real parameters such that $\max_{x\in \X} \abs{g_1(x) + i \cdot g_2(x) - \log(f(x))} < \epsilon$,
where $c(\epsilon, m, W_{\max}, f_{\min}) {\equiv} O\!\left(\ln^2\left(\frac{m}{\epsilon} \ln\!\left(\frac{W_{\max}}{f_{\min}}\right)\right) + \ln\!\left(\frac{1}{\epsilon}\right)\!\sqrt{\frac{1}{\epsilon}}\right)$.
\end{theorem}

The proof of Theorem~\ref{th:main}, which is given in full in app.~\ref{app:proof}, is based on two steps. First, we show that AC with non-negative parameters and inputs can be exactly reconstructed with NN with real parameters and softplus activation functions. In this simple case, for any intermediate values $x_1,x_2 \geq 0$, we can set $o_i = \log(x_i)$(where 0 is mapped to the right-side limit of $-\infty$), and then multiplication becomes summation, i.e., $\log(x_1 \cdot x_2) = o_1 + o_2$. For summation, softplus activations arise naturally:
\begin{align}
\log(x_1 {+} x_2) &= \log(e^{o_1} {+} e^{o_2}) = o_1 {+} \log\left(1 {+} e^{o_2 {-} o_1} \right) \nonumber \\
&= o_1 {+} \softplus(o_2 {-} o_1). \label{eq:log_sum_exp_to_softplus}
\end{align}
For log-space summation of $n$ inputs, we can decompose it as a binary tree, which gives the $\log(m)$ correction to the depth of the network.
Second, we reduce the complex case to the non-negative case plus a finite number of smooth operations, which can be approximated efficiently by employing various techniques.
Since only a finite number of operations requires approximation, it results in the additive term $c(\epsilon,m,W_{\max},f_{\min})$, which is merely logarithmic in the number of edges of the AC, and double logarithmic with respect to the magnitudes of the weights and the WF amplitudes.
These weak dependencies of the target AC result in practically arbitrary precision.
The immediate implication of Theorem~\ref{th:main} is that NQS can simulate TNS at least as efficiently as their TN representation:
\begin{corollary}\label{co:tns_as_nqs}
For any tensor network quantum state with a contraction scheme of run-time $k$, and at most $b$ bits of precision in computations and parameters, there exists a neural network that approximate it with a maximal error of $\epsilon$ and of run-time (number of edges) $O\left(k + \ln^2\left(\frac{kb}{\epsilon}\right) + \ln\left(\frac{1}{\epsilon}\right)\sqrt{\frac{1}{\epsilon}}\right)$.
\end{corollary}

\begin{figure*}[!ht]
    \centering
    \includegraphics[width=\linewidth,keepaspectratio]{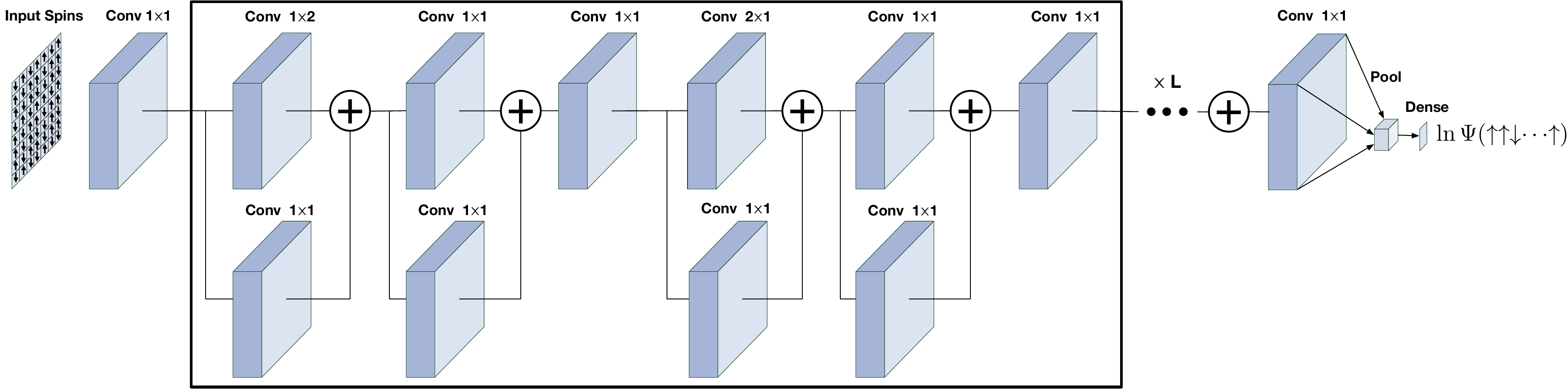}
    \caption{\label{fig:volume_law_convnet} An illustration of a convolutional neural network~(ConvNet), according to Cor.~\ref{co:nqs_exp}, that is capable of representing two-dimensional quantum spin ($d=2$) states with volume-law entanglement entropy that a PEPS model cannot represent unless it employs exponential (in number of sites) parameters. The ConvNet is made up of a sequence of $L$ blocks. Each block has two spatial (either $2{\times}1$ or $1{\times}2$ window size) convolutional layers, eight (in general, $2\log_2(2d)$) local convolutional layers ($1{\times}1$ windows) and residual connections between them. The residual connections are a direct result of the $o_1 + \mathrm{softplus}(o_2-o_1)$ construction found in Eq.~\ref{eq:log_sum_exp_to_softplus}. There are no pooling layers except in the next to last layer, in which a global average pooling operation is applied. The network ends with a dense layer reducing the dimension to a scalar that represents the log-amplitude of the quantum state. If a system employs a $q{\times}q$ grid ($N=q^2$), then a ConvNet with $L=\nicefrac{q}{2}$ blocks can be used to represent some volume-law states, which PEPS cannot represent efficiently. Similar constructions can be used for 1D~--~separating NN from both MPS and MERA~--~as well as for higher spatial dimensions and for higher local dimensions $d > 2$.}
\end{figure*}

For the specific case of MPS, Cor.~\ref{co:tns_as_nqs} translates to the following:
\begin{corollary}\label{co:mps_rep}
For any MPS over $N$ sites, each of local dimension $d$, with bond dimension $\chi$, and fixed $b$ bits of precision, there exists a neural network of depth $l$ consisting of $m$ edges that approximates its contraction algorithm up to $\epsilon$, where $l$ and $m$ depend on the contraction scheme\footnote{The bounds on the two contraction schemes serve to highlight different characteristics. By using the sequential scheme, it is demonstrated that NN can approximate MPS with the same optimal runtime when not accounting for parallelization. By employing the parallel scheme, it demonstrates that a logarithmic depth is sufficient for approximating MPS, while also better utilizing parallel execution as supported by modern GPUs.}:
\vspace{-3mm}
\begin{enumerate}
\setlength{\itemindent}{-.3in}
\setlength{\itemsep}{0pt}%
    \setlength{\parskip}{0pt}%
    \item Sequential: $l = \tilde{O}\left(N +\sqrt{\nicefrac{1}{\epsilon}}\right)$ and $m = \tilde{O}\left(N d \chi^2 + \sqrt{\nicefrac{1}{\epsilon}}\right)$.
    \item Parallel: $\tilde{O}(\ln N + \sqrt{\nicefrac{1}{\epsilon}})$ and $m = \tilde{O}\left(N (d + \chi)\chi^2 + \sqrt{\nicefrac{1}{\epsilon}}\right)$.
\end{enumerate}
\vspace{-3mm}
where $\tilde{O}$ denotes big-O while ignoring logarithmic factors.
\end{corollary}

In turn, this result also allows to use previously established rigorous results on MPS to directly quantify the expressive power of NQS on special classes of quantum systems.
For example, Hastings famously established an area-law entanglement for the gapped ground state of one-dimensional systems \cite{hastings_area_2007} that directly translates into an efficient approximation by MPS \cite{hastings_area_2007,arad_area_2013,schuch_matrix_2017,dalzell_locally_2019}.
Our result in \ref{co:mps_rep}, in connection with the bound established in \cite{hastings_area_2007} implies:   

\begin{corollary}\label{co:1d_gapped}
    Consider a 1D Hamiltonian $H$ defined on $N$ qudits of finite local dimension $d$, and with a non-vanishing spectral gap $\Delta$. The ground state of a $H$ can be written as a deep neural network of depth ${l = O(\ln N {+} \sqrt{\nicefrac{1}{\epsilon}})}$ and ${m {=} O\left(\mathrm{poly}(N,1/\epsilon)\right)}$ edges. 
\end{corollary}

Though we have established a strictly inclusive relationship, we show that the reverse is not true, that is, there are NQS that cannot be efficiently reproduced by the most commonly used variational TNS.:
\begin{corollary}\label{co:nqs_exp}
    There exist quantum states that can be represented by NN with parameters and runtime polynomial in the number of sites, that MPS and PEPS cannot represent efficiently unless they use exponentially many parameters.
\end{corollary}
The proof is based on prior results~\citep{levine2019quantum,sharir2018expressive} that used convolutional AC as indirect analogs to convolutional NN, and showed that convolutional AC can represent some volume-law states, which MPS and PEPS cannot represent efficiently.
Using Theorem~\ref{th:main} we can close this theoretical gap and transfer these results to real-world NN.
See Fig.~\ref{fig:volume_law_convnet} for an illustration.
Cor.~\ref{co:nqs_exp} leaves open the possibility of other novel geometries for TNS that could be efficient evaluated while supporting volume-law states.
Nevertheless, using Theorem~\ref{th:main}, even such TNS could be represented by NQS.
Overall, we have then established the representability diagram of Fig.~\ref{fig:diagram}. 

\paragraph*{Discussion --}
In this work we have introduced a general mapping between tensor networks and deep artificial neural networks. 
This mapping allows to directly connect two of the most important classes of parametric representations of high dimensional functions,
and allows to establish a representation diagram of modern variational many-body variational quantum states.
Moreover, while it is beyond the scope of the main text, our results could be extended to support approximated contraction schemes as well (See App.~\ref{app:approximated_contractions}).

We expect that our mapping will be especially useful to establish further rigorous representation results on neural-network based quantum 
states, using the well-developed theory of tensor-network representations.
Indeed, our analysis is limited in scope to TNS that support efficient amplitude evaluation, i.e., $\langle s | \psi\rangle$, which do cover all planar TNS that are most commonly adopted in practical computations.
However, our work also lays the important foundations for studying the relation to TNS that support efficient computation of expectation values, for a restricted set of observables, even if they lack the ability to efficiently compute (or approximate well) $\langle s | \psi\rangle$, e.g., the MERA family of ansatzs~\citep{PhysRevLett.112.240502,evenbly2014scaling}.

Furthermore, the kind of neural-network architectures
and connectivity patterns resulting from our mapping might also inspire new practical applications inspired by successful tensor-network ideas. 
Along the same lines, our mapping can also help clarify in what circumstances gradient-based optimization
strategies, ubiquitous in machine learning, are to be preferred over successful alternated optimization strategies instead commonly adopted for tensor networks. 

\begin{acknowledgments}
This research was supported by the ERC (European Research
Council) and the ISF (Israel Science Foundation).
\end{acknowledgments}

\vspace{0.5cm}

\section*{References}

\bibliography{refs}

\begin{thebibliography}{49}%
\makeatletter
\providecommand \@ifxundefined [1]{%
 \@ifx{#1\undefined}
}%
\providecommand \@ifnum [1]{%
 \ifnum #1\expandafter \@firstoftwo
 \else \expandafter \@secondoftwo
 \fi
}%
\providecommand \@ifx [1]{%
 \ifx #1\expandafter \@firstoftwo
 \else \expandafter \@secondoftwo
 \fi
}%
\providecommand \natexlab [1]{#1}%
\providecommand \enquote  [1]{``#1''}%
\providecommand \bibnamefont  [1]{#1}%
\providecommand \bibfnamefont [1]{#1}%
\providecommand \citenamefont [1]{#1}%
\providecommand \href@noop [0]{\@secondoftwo}%
\providecommand \href [0]{\begingroup \@sanitize@url \@href}%
\providecommand \@href[1]{\@@startlink{#1}\@@href}%
\providecommand \@@href[1]{\endgroup#1\@@endlink}%
\providecommand \@sanitize@url [0]{\catcode `\\12\catcode `\$12\catcode
  `\&12\catcode `\#12\catcode `\^12\catcode `\_12\catcode `\%12\relax}%
\providecommand \@@startlink[1]{}%
\providecommand \@@endlink[0]{}%
\providecommand \url  [0]{\begingroup\@sanitize@url \@url }%
\providecommand \@url [1]{\endgroup\@href {#1}{\urlprefix }}%
\providecommand \urlprefix  [0]{URL }%
\providecommand \Eprint [0]{\href }%
\providecommand \doibase [0]{https://doi.org/}%
\providecommand \selectlanguage [0]{\@gobble}%
\providecommand \bibinfo  [0]{\@secondoftwo}%
\providecommand \bibfield  [0]{\@secondoftwo}%
\providecommand \translation [1]{[#1]}%
\providecommand \BibitemOpen [0]{}%
\providecommand \bibitemStop [0]{}%
\providecommand \bibitemNoStop [0]{.\EOS\space}%
\providecommand \EOS [0]{\spacefactor3000\relax}%
\providecommand \BibitemShut  [1]{\csname bibitem#1\endcsname}%
\let\auto@bib@innerbib\@empty
\bibitem [{\citenamefont {Goodfellow}\ \emph {et~al.}(2016)\citenamefont
  {Goodfellow}, \citenamefont {Bengio},\ and\ \citenamefont
  {Courville}}]{goodfellow_deep_2016}%
  \BibitemOpen
  \bibfield  {author} {\bibinfo {author} {\bibfnamefont {I.}~\bibnamefont
  {Goodfellow}}, \bibinfo {author} {\bibfnamefont {Y.}~\bibnamefont {Bengio}},\
  and\ \bibinfo {author} {\bibfnamefont {A.}~\bibnamefont {Courville}},\
  }\href@noop {} {\emph {\bibinfo {title} {Deep Learning}}}\ (\bibinfo
  {publisher} {The MIT Press},\ \bibinfo {address} {Cambridge, Massachusetts},\
  \bibinfo {year} {2016})\BibitemShut {NoStop}%
\bibitem [{\citenamefont {Carleo}\ \emph {et~al.}(2019)\citenamefont {Carleo},
  \citenamefont {Cirac}, \citenamefont {Cranmer}, \citenamefont {Daudet},
  \citenamefont {Schuld}, \citenamefont {Tishby}, \citenamefont
  {Vogt-Maranto},\ and\ \citenamefont {Zdeborov\'a}}]{carleo_machine_2019}%
  \BibitemOpen
  \bibfield  {author} {\bibinfo {author} {\bibfnamefont {G.}~\bibnamefont
  {Carleo}}, \bibinfo {author} {\bibfnamefont {I.}~\bibnamefont {Cirac}},
  \bibinfo {author} {\bibfnamefont {K.}~\bibnamefont {Cranmer}}, \bibinfo
  {author} {\bibfnamefont {L.}~\bibnamefont {Daudet}}, \bibinfo {author}
  {\bibfnamefont {M.}~\bibnamefont {Schuld}}, \bibinfo {author} {\bibfnamefont
  {N.}~\bibnamefont {Tishby}}, \bibinfo {author} {\bibfnamefont
  {L.}~\bibnamefont {Vogt-Maranto}},\ and\ \bibinfo {author} {\bibfnamefont
  {L.}~\bibnamefont {Zdeborov\'a}},\ }\bibfield  {title} {\bibinfo {title}
  {Machine learning and the physical sciences},\ }\href
  {https://doi.org/10.1103/RevModPhys.91.045002} {\bibfield  {journal}
  {\bibinfo  {journal} {Rev. Mod. Phys.}\ }\textbf {\bibinfo {volume} {91}},\
  \bibinfo {pages} {045002} (\bibinfo {year} {2019})}\BibitemShut {NoStop}%
\bibitem [{\citenamefont {White}(1992)}]{white1992density}%
  \BibitemOpen
  \bibfield  {author} {\bibinfo {author} {\bibfnamefont {S.~R.}\ \bibnamefont
  {White}},\ }\bibfield  {title} {\bibinfo {title} {Density matrix formulation
  for quantum renormalization groups},\ }\href@noop {} {\bibfield  {journal}
  {\bibinfo  {journal} {Physical review letters}\ }\textbf {\bibinfo {volume}
  {69}},\ \bibinfo {pages} {2863} (\bibinfo {year} {1992})}\BibitemShut
  {NoStop}%
\bibitem [{\citenamefont {Schollw{\"o}ck}(2011)}]{schollwock2011density}%
  \BibitemOpen
  \bibfield  {author} {\bibinfo {author} {\bibfnamefont {U.}~\bibnamefont
  {Schollw{\"o}ck}},\ }\bibfield  {title} {\bibinfo {title} {The density-matrix
  renormalization group in the age of matrix product states},\ }\href@noop {}
  {\bibfield  {journal} {\bibinfo  {journal} {Annals of Physics}\ }\textbf
  {\bibinfo {volume} {326}},\ \bibinfo {pages} {96} (\bibinfo {year}
  {2011})}\BibitemShut {NoStop}%
\bibitem [{\citenamefont {Orús}(2019)}]{orus_tensor_2019}%
  \BibitemOpen
  \bibfield  {author} {\bibinfo {author} {\bibfnamefont {R.}~\bibnamefont
  {Orús}},\ }\bibfield  {title} {\bibinfo {title} {Tensor networks for complex
  quantum systems},\ }\href {https://doi.org/10.1038/s42254-019-0086-7}
  {\bibfield  {journal} {\bibinfo  {journal} {Nature Reviews Physics}\ }\textbf
  {\bibinfo {volume} {1}},\ \bibinfo {pages} {538} (\bibinfo {year} {2019})},\
  \bibinfo {note} {number: 9 Publisher: Nature Publishing Group}\BibitemShut
  {NoStop}%
\bibitem [{\citenamefont {Verstraete}\ \emph {et~al.}(2008)\citenamefont
  {Verstraete}, \citenamefont {Murg},\ and\ \citenamefont
  {Cirac}}]{verstraete2008matrix}%
  \BibitemOpen
  \bibfield  {author} {\bibinfo {author} {\bibfnamefont {F.}~\bibnamefont
  {Verstraete}}, \bibinfo {author} {\bibfnamefont {V.}~\bibnamefont {Murg}},\
  and\ \bibinfo {author} {\bibfnamefont {J.~I.}\ \bibnamefont {Cirac}},\
  }\bibfield  {title} {\bibinfo {title} {Matrix product states, projected
  entangled pair states, and variational renormalization group methods for
  quantum spin systems},\ }\href@noop {} {\bibfield  {journal} {\bibinfo
  {journal} {Advances in Physics}\ }\textbf {\bibinfo {volume} {57}},\ \bibinfo
  {pages} {143} (\bibinfo {year} {2008})}\BibitemShut {NoStop}%
\bibitem [{\citenamefont {Cirac}\ \emph {et~al.}(2020)\citenamefont {Cirac},
  \citenamefont {Perez-Garcia}, \citenamefont {Schuch},\ and\ \citenamefont
  {Verstraete}}]{cirac_matrix_2020}%
  \BibitemOpen
  \bibfield  {author} {\bibinfo {author} {\bibfnamefont {I.}~\bibnamefont
  {Cirac}}, \bibinfo {author} {\bibfnamefont {D.}~\bibnamefont {Perez-Garcia}},
  \bibinfo {author} {\bibfnamefont {N.}~\bibnamefont {Schuch}},\ and\ \bibinfo
  {author} {\bibfnamefont {F.}~\bibnamefont {Verstraete}},\ }\bibfield  {title}
  {\bibinfo {title} {Matrix {Product} {States} and {Projected} {Entangled}
  {Pair} {States}: {Concepts}, {Symmetries}, and {Theorems}},\ }\href
  {http://arxiv.org/abs/2011.12127} {\bibfield  {journal} {\bibinfo  {journal}
  {arXiv:2011.12127 [cond-mat, physics:hep-th, physics:quant-ph]}\ } (\bibinfo
  {year} {2020})},\ \bibinfo {note} {arXiv: 2011.12127}\BibitemShut {NoStop}%
\bibitem [{\citenamefont {Carleo}\ and\ \citenamefont
  {Troyer}(2017)}]{carleo2017solving}%
  \BibitemOpen
  \bibfield  {author} {\bibinfo {author} {\bibfnamefont {G.}~\bibnamefont
  {Carleo}}\ and\ \bibinfo {author} {\bibfnamefont {M.}~\bibnamefont
  {Troyer}},\ }\bibfield  {title} {\bibinfo {title} {Solving the quantum
  many-body problem with artificial neural networks},\ }\href@noop {}
  {\bibfield  {journal} {\bibinfo  {journal} {Science}\ }\textbf {\bibinfo
  {volume} {355}},\ \bibinfo {pages} {602} (\bibinfo {year}
  {2017})}\BibitemShut {NoStop}%
\bibitem [{\citenamefont {Cybenko}(1989)}]{Cybenko:1989fm}%
  \BibitemOpen
  \bibfield  {author} {\bibinfo {author} {\bibfnamefont {G.}~\bibnamefont
  {Cybenko}},\ }\bibfield  {title} {\bibinfo {title} {{Approximation by
  superpositions of a sigmoidal function}},\ }\href@noop {} {\bibfield
  {journal} {\bibinfo  {journal} {Mathematics of Control, Signals and Systems}\
  }\textbf {\bibinfo {volume} {2}},\ \bibinfo {pages} {303} (\bibinfo {year}
  {1989})}\BibitemShut {NoStop}%
\bibitem [{\citenamefont {Carleo}\ \emph {et~al.}(2018)\citenamefont {Carleo},
  \citenamefont {Nomura},\ and\ \citenamefont
  {Imada}}]{carleo_constructing_2018}%
  \BibitemOpen
  \bibfield  {author} {\bibinfo {author} {\bibfnamefont {G.}~\bibnamefont
  {Carleo}}, \bibinfo {author} {\bibfnamefont {Y.}~\bibnamefont {Nomura}},\
  and\ \bibinfo {author} {\bibfnamefont {M.}~\bibnamefont {Imada}},\ }\bibfield
   {title} {\bibinfo {title} {Constructing exact representations of quantum
  many-body systems with deep neural networks},\ }\href
  {https://doi.org/10.1038/s41467-018-07520-3} {\bibfield  {journal} {\bibinfo
  {journal} {Nature Communications}\ }\textbf {\bibinfo {volume} {9}},\
  \bibinfo {pages} {5322} (\bibinfo {year} {2018})}\BibitemShut {NoStop}%
\bibitem [{\citenamefont {Gao}\ and\ \citenamefont
  {Duan}(2017)}]{gao2017efficient}%
  \BibitemOpen
  \bibfield  {author} {\bibinfo {author} {\bibfnamefont {X.}~\bibnamefont
  {Gao}}\ and\ \bibinfo {author} {\bibfnamefont {L.-M.}\ \bibnamefont {Duan}},\
  }\bibfield  {title} {\bibinfo {title} {Efficient representation of quantum
  many-body states with deep neural networks},\ }\href@noop {} {\bibfield
  {journal} {\bibinfo  {journal} {Nature communications}\ }\textbf {\bibinfo
  {volume} {8}},\ \bibinfo {pages} {662} (\bibinfo {year} {2017})}\BibitemShut
  {NoStop}%
\bibitem [{\citenamefont {Deng}\ \emph
  {et~al.}(2017{\natexlab{a}})\citenamefont {Deng}, \citenamefont {Li},\ and\
  \citenamefont {Das~Sarma}}]{deng_machine_2017}%
  \BibitemOpen
  \bibfield  {author} {\bibinfo {author} {\bibfnamefont {D.-L.}\ \bibnamefont
  {Deng}}, \bibinfo {author} {\bibfnamefont {X.}~\bibnamefont {Li}},\ and\
  \bibinfo {author} {\bibfnamefont {S.}~\bibnamefont {Das~Sarma}},\ }\bibfield
  {title} {\bibinfo {title} {Machine learning topological states},\ }\href
  {https://doi.org/10.1103/PhysRevB.96.195145} {\bibfield  {journal} {\bibinfo
  {journal} {Physical Review B}\ }\textbf {\bibinfo {volume} {96}},\ \bibinfo
  {pages} {195145} (\bibinfo {year} {2017}{\natexlab{a}})}\BibitemShut
  {NoStop}%
\bibitem [{\citenamefont {Kaubruegger}\ \emph {et~al.}(2018)\citenamefont
  {Kaubruegger}, \citenamefont {Pastori},\ and\ \citenamefont
  {Budich}}]{kaubruegger_chiral_2018}%
  \BibitemOpen
  \bibfield  {author} {\bibinfo {author} {\bibfnamefont {R.}~\bibnamefont
  {Kaubruegger}}, \bibinfo {author} {\bibfnamefont {L.}~\bibnamefont
  {Pastori}},\ and\ \bibinfo {author} {\bibfnamefont {J.~C.}\ \bibnamefont
  {Budich}},\ }\bibfield  {title} {\bibinfo {title} {Chiral topological phases
  from artificial neural networks},\ }\href
  {https://doi.org/10.1103/PhysRevB.97.195136} {\bibfield  {journal} {\bibinfo
  {journal} {Physical Review B}\ }\textbf {\bibinfo {volume} {97}},\ \bibinfo
  {pages} {195136} (\bibinfo {year} {2018})}\BibitemShut {NoStop}%
\bibitem [{\citenamefont {Glasser}\ \emph
  {et~al.}(2018{\natexlab{a}})\citenamefont {Glasser}, \citenamefont
  {Pancotti}, \citenamefont {August}, \citenamefont {Rodriguez},\ and\
  \citenamefont {Cirac}}]{glasser_neural-network_2018}%
  \BibitemOpen
  \bibfield  {author} {\bibinfo {author} {\bibfnamefont {I.}~\bibnamefont
  {Glasser}}, \bibinfo {author} {\bibfnamefont {N.}~\bibnamefont {Pancotti}},
  \bibinfo {author} {\bibfnamefont {M.}~\bibnamefont {August}}, \bibinfo
  {author} {\bibfnamefont {I.~D.}\ \bibnamefont {Rodriguez}},\ and\ \bibinfo
  {author} {\bibfnamefont {J.~I.}\ \bibnamefont {Cirac}},\ }\bibfield  {title}
  {\bibinfo {title} {Neural-{Network} {Quantum} {States}, {String}-{Bond}
  {States}, and {Chiral} {Topological} {States}},\ }\href
  {https://doi.org/10.1103/PhysRevX.8.011006} {\bibfield  {journal} {\bibinfo
  {journal} {Physical Review X}\ }\textbf {\bibinfo {volume} {8}},\ \bibinfo
  {pages} {011006} (\bibinfo {year} {2018}{\natexlab{a}})}\BibitemShut
  {NoStop}%
\bibitem [{\citenamefont {Lu}\ \emph {et~al.}(2019)\citenamefont {Lu},
  \citenamefont {Gao},\ and\ \citenamefont {Duan}}]{lu_efficient_2019}%
  \BibitemOpen
  \bibfield  {author} {\bibinfo {author} {\bibfnamefont {S.}~\bibnamefont
  {Lu}}, \bibinfo {author} {\bibfnamefont {X.}~\bibnamefont {Gao}},\ and\
  \bibinfo {author} {\bibfnamefont {L.-M.}\ \bibnamefont {Duan}},\ }\bibfield
  {title} {\bibinfo {title} {Efficient representation of topologically ordered
  states with restricted {Boltzmann} machines},\ }\href
  {https://doi.org/10.1103/PhysRevB.99.155136} {\bibfield  {journal} {\bibinfo
  {journal} {Physical Review B}\ }\textbf {\bibinfo {volume} {99}},\ \bibinfo
  {pages} {155136} (\bibinfo {year} {2019})}\BibitemShut {NoStop}%
\bibitem [{\citenamefont {Glasser}\ \emph
  {et~al.}(2018{\natexlab{b}})\citenamefont {Glasser}, \citenamefont
  {Pancotti}, \citenamefont {August}, \citenamefont {Rodriguez},\ and\
  \citenamefont {Cirac}}]{PhysRevX.8.011006}%
  \BibitemOpen
  \bibfield  {author} {\bibinfo {author} {\bibfnamefont {I.}~\bibnamefont
  {Glasser}}, \bibinfo {author} {\bibfnamefont {N.}~\bibnamefont {Pancotti}},
  \bibinfo {author} {\bibfnamefont {M.}~\bibnamefont {August}}, \bibinfo
  {author} {\bibfnamefont {I.~D.}\ \bibnamefont {Rodriguez}},\ and\ \bibinfo
  {author} {\bibfnamefont {J.~I.}\ \bibnamefont {Cirac}},\ }\bibfield  {title}
  {\bibinfo {title} {Neural-network quantum states, string-bond states, and
  chiral topological states},\ }\href
  {https://doi.org/10.1103/PhysRevX.8.011006} {\bibfield  {journal} {\bibinfo
  {journal} {Phys. Rev. X}\ }\textbf {\bibinfo {volume} {8}},\ \bibinfo {pages}
  {011006} (\bibinfo {year} {2018}{\natexlab{b}})}\BibitemShut {NoStop}%
\bibitem [{\citenamefont {Choo}\ \emph {et~al.}(2019)\citenamefont {Choo},
  \citenamefont {Neupert},\ and\ \citenamefont
  {Carleo}}]{choo_two-dimensional_2019}%
  \BibitemOpen
  \bibfield  {author} {\bibinfo {author} {\bibfnamefont {K.}~\bibnamefont
  {Choo}}, \bibinfo {author} {\bibfnamefont {T.}~\bibnamefont {Neupert}},\ and\
  \bibinfo {author} {\bibfnamefont {G.}~\bibnamefont {Carleo}},\ }\bibfield
  {title} {\bibinfo {title} {Two-dimensional frustrated
  \$\{{J}\}\_\{1\}{\textbackslash}text\{{\textbackslash}ensuremath\{-\}\}\{{J}\}\_\{2\}\$
  model studied with neural network quantum states},\ }\href
  {https://doi.org/10.1103/PhysRevB.100.125124} {\bibfield  {journal} {\bibinfo
   {journal} {Physical Review B}\ }\textbf {\bibinfo {volume} {100}},\ \bibinfo
  {pages} {125124} (\bibinfo {year} {2019})}\BibitemShut {NoStop}%
\bibitem [{\citenamefont {Sharir}\ \emph {et~al.}(2020)\citenamefont {Sharir},
  \citenamefont {Levine}, \citenamefont {Wies}, \citenamefont {Carleo},\ and\
  \citenamefont {Shashua}}]{sharir_deep_2020}%
  \BibitemOpen
  \bibfield  {author} {\bibinfo {author} {\bibfnamefont {O.}~\bibnamefont
  {Sharir}}, \bibinfo {author} {\bibfnamefont {Y.}~\bibnamefont {Levine}},
  \bibinfo {author} {\bibfnamefont {N.}~\bibnamefont {Wies}}, \bibinfo {author}
  {\bibfnamefont {G.}~\bibnamefont {Carleo}},\ and\ \bibinfo {author}
  {\bibfnamefont {A.}~\bibnamefont {Shashua}},\ }\bibfield  {title} {\bibinfo
  {title} {Deep {Autoregressive} {Models} for the {Efficient} {Variational}
  {Simulation} of {Many}-{Body} {Quantum} {Systems}},\ }\href
  {https://doi.org/10.1103/PhysRevLett.124.020503} {\bibfield  {journal}
  {\bibinfo  {journal} {Physical Review Letters}\ }\textbf {\bibinfo {volume}
  {124}},\ \bibinfo {pages} {020503} (\bibinfo {year} {2020})},\ \bibinfo
  {note} {publisher: American Physical Society}\BibitemShut {NoStop}%
\bibitem [{\citenamefont {Schmitt}\ and\ \citenamefont
  {Heyl}(2020)}]{schmitt_quantum_2020}%
  \BibitemOpen
  \bibfield  {author} {\bibinfo {author} {\bibfnamefont {M.}~\bibnamefont
  {Schmitt}}\ and\ \bibinfo {author} {\bibfnamefont {M.}~\bibnamefont {Heyl}},\
  }\bibfield  {title} {\bibinfo {title} {Quantum {Many}-{Body} {Dynamics} in
  {Two} {Dimensions} with {Artificial} {Neural} {Networks}},\ }\href
  {https://doi.org/10.1103/PhysRevLett.125.100503} {\bibfield  {journal}
  {\bibinfo  {journal} {Physical Review Letters}\ }\textbf {\bibinfo {volume}
  {125}},\ \bibinfo {pages} {100503} (\bibinfo {year} {2020})},\ \bibinfo
  {note} {publisher: American Physical Society}\BibitemShut {NoStop}%
\bibitem [{\citenamefont {Hibat-Allah}\ \emph {et~al.}(2020)\citenamefont
  {Hibat-Allah}, \citenamefont {Ganahl}, \citenamefont {Hayward}, \citenamefont
  {Melko},\ and\ \citenamefont {Carrasquilla}}]{hibat-allah_recurrent_2020}%
  \BibitemOpen
  \bibfield  {author} {\bibinfo {author} {\bibfnamefont {M.}~\bibnamefont
  {Hibat-Allah}}, \bibinfo {author} {\bibfnamefont {M.}~\bibnamefont {Ganahl}},
  \bibinfo {author} {\bibfnamefont {L.~E.}\ \bibnamefont {Hayward}}, \bibinfo
  {author} {\bibfnamefont {R.~G.}\ \bibnamefont {Melko}},\ and\ \bibinfo
  {author} {\bibfnamefont {J.}~\bibnamefont {Carrasquilla}},\ }\bibfield
  {title} {\bibinfo {title} {Recurrent neural network wave functions},\ }\href
  {https://doi.org/10.1103/PhysRevResearch.2.023358} {\bibfield  {journal}
  {\bibinfo  {journal} {Physical Review Research}\ }\textbf {\bibinfo {volume}
  {2}},\ \bibinfo {pages} {023358} (\bibinfo {year} {2020})},\ \bibinfo {note}
  {publisher: American Physical Society}\BibitemShut {NoStop}%
\bibitem [{\citenamefont {Torlai}\ \emph {et~al.}(2018)\citenamefont {Torlai},
  \citenamefont {Mazzola}, \citenamefont {Carrasquilla}, \citenamefont
  {Troyer}, \citenamefont {Melko},\ and\ \citenamefont
  {Carleo}}]{torlai_neural-network_2018}%
  \BibitemOpen
  \bibfield  {author} {\bibinfo {author} {\bibfnamefont {G.}~\bibnamefont
  {Torlai}}, \bibinfo {author} {\bibfnamefont {G.}~\bibnamefont {Mazzola}},
  \bibinfo {author} {\bibfnamefont {J.}~\bibnamefont {Carrasquilla}}, \bibinfo
  {author} {\bibfnamefont {M.}~\bibnamefont {Troyer}}, \bibinfo {author}
  {\bibfnamefont {R.}~\bibnamefont {Melko}},\ and\ \bibinfo {author}
  {\bibfnamefont {G.}~\bibnamefont {Carleo}},\ }\bibfield  {title} {\bibinfo
  {title} {Neural-network quantum state tomography},\ }\href
  {https://doi.org/10.1038/s41567-018-0048-5} {\bibfield  {journal} {\bibinfo
  {journal} {Nature Physics}\ }\textbf {\bibinfo {volume} {14}},\ \bibinfo
  {pages} {447} (\bibinfo {year} {2018})}\BibitemShut {NoStop}%
\bibitem [{\citenamefont {Pfau}\ \emph {et~al.}(2020)\citenamefont {Pfau},
  \citenamefont {Spencer}, \citenamefont {Matthews},\ and\ \citenamefont
  {Foulkes}}]{pfau_ab_2020}%
  \BibitemOpen
  \bibfield  {author} {\bibinfo {author} {\bibfnamefont {D.}~\bibnamefont
  {Pfau}}, \bibinfo {author} {\bibfnamefont {J.~S.}\ \bibnamefont {Spencer}},
  \bibinfo {author} {\bibfnamefont {A.~G. D.~G.}\ \bibnamefont {Matthews}},\
  and\ \bibinfo {author} {\bibfnamefont {W.~M.~C.}\ \bibnamefont {Foulkes}},\
  }\bibfield  {title} {\bibinfo {title} {Ab initio solution of the
  many-electron {Schr}{\textbackslash}"odinger equation with deep neural
  networks},\ }\href {https://doi.org/10.1103/PhysRevResearch.2.033429}
  {\bibfield  {journal} {\bibinfo  {journal} {Physical Review Research}\
  }\textbf {\bibinfo {volume} {2}},\ \bibinfo {pages} {033429} (\bibinfo {year}
  {2020})},\ \bibinfo {note} {publisher: American Physical Society}\BibitemShut
  {NoStop}%
\bibitem [{\citenamefont {Hermann}\ \emph {et~al.}(2020)\citenamefont
  {Hermann}, \citenamefont {Schätzle},\ and\ \citenamefont
  {Noé}}]{hermann_deep-neural-network_2020}%
  \BibitemOpen
  \bibfield  {author} {\bibinfo {author} {\bibfnamefont {J.}~\bibnamefont
  {Hermann}}, \bibinfo {author} {\bibfnamefont {Z.}~\bibnamefont {Schätzle}},\
  and\ \bibinfo {author} {\bibfnamefont {F.}~\bibnamefont {Noé}},\ }\bibfield
  {title} {\bibinfo {title} {Deep-neural-network solution of the electronic
  {Schrödinger} equation},\ }\href {https://doi.org/10.1038/s41557-020-0544-y}
  {\bibfield  {journal} {\bibinfo  {journal} {Nature Chemistry}\ }\textbf
  {\bibinfo {volume} {12}},\ \bibinfo {pages} {891} (\bibinfo {year} {2020})},\
  \bibinfo {note} {number: 10 Publisher: Nature Publishing Group}\BibitemShut
  {NoStop}%
\bibitem [{\citenamefont {Choo}\ \emph {et~al.}(2020)\citenamefont {Choo},
  \citenamefont {Mezzacapo},\ and\ \citenamefont
  {Carleo}}]{choo_fermionic_2020}%
  \BibitemOpen
  \bibfield  {author} {\bibinfo {author} {\bibfnamefont {K.}~\bibnamefont
  {Choo}}, \bibinfo {author} {\bibfnamefont {A.}~\bibnamefont {Mezzacapo}},\
  and\ \bibinfo {author} {\bibfnamefont {G.}~\bibnamefont {Carleo}},\
  }\bibfield  {title} {\bibinfo {title} {Fermionic neural-network states for
  ab-initio electronic structure},\ }\href
  {https://doi.org/10.1038/s41467-020-15724-9} {\bibfield  {journal} {\bibinfo
  {journal} {Nature Communications}\ }\textbf {\bibinfo {volume} {11}},\
  \bibinfo {pages} {2368} (\bibinfo {year} {2020})},\ \bibinfo {note} {number:
  1 Publisher: Nature Publishing Group}\BibitemShut {NoStop}%
\bibitem [{\citenamefont {Verstraete}\ and\ \citenamefont
  {Cirac}(2004)}]{verstraete2004renormalization}%
  \BibitemOpen
  \bibfield  {author} {\bibinfo {author} {\bibfnamefont {F.}~\bibnamefont
  {Verstraete}}\ and\ \bibinfo {author} {\bibfnamefont {J.~I.}\ \bibnamefont
  {Cirac}},\ }\bibfield  {title} {\bibinfo {title} {Renormalization algorithms
  for quantum-many body systems in two and higher dimensions},\ }\href@noop {}
  {\bibfield  {journal} {\bibinfo  {journal} {arXiv preprint cond-mat/0407066}\
  } (\bibinfo {year} {2004})}\BibitemShut {NoStop}%
\bibitem [{\citenamefont {Schuch}\ \emph {et~al.}(2007)\citenamefont {Schuch},
  \citenamefont {Wolf}, \citenamefont {Verstraete},\ and\ \citenamefont
  {Cirac}}]{schuch_computational_2007}%
  \BibitemOpen
  \bibfield  {author} {\bibinfo {author} {\bibfnamefont {N.}~\bibnamefont
  {Schuch}}, \bibinfo {author} {\bibfnamefont {M.~M.}\ \bibnamefont {Wolf}},
  \bibinfo {author} {\bibfnamefont {F.}~\bibnamefont {Verstraete}},\ and\
  \bibinfo {author} {\bibfnamefont {J.~I.}\ \bibnamefont {Cirac}},\ }\bibfield
  {title} {\bibinfo {title} {Computational {Complexity} of {Projected}
  {Entangled} {Pair} {States}},\ }\href
  {https://doi.org/10.1103/PhysRevLett.98.140506} {\bibfield  {journal}
  {\bibinfo  {journal} {Physical Review Letters}\ }\textbf {\bibinfo {volume}
  {98}},\ \bibinfo {pages} {140506} (\bibinfo {year} {2007})},\ \bibinfo {note}
  {publisher: American Physical Society}\BibitemShut {NoStop}%
\bibitem [{\citenamefont {Haferkamp}\ \emph {et~al.}(2020)\citenamefont
  {Haferkamp}, \citenamefont {Hangleiter}, \citenamefont {Eisert},\ and\
  \citenamefont {Gluza}}]{haferkamp_contracting_2020}%
  \BibitemOpen
  \bibfield  {author} {\bibinfo {author} {\bibfnamefont {J.}~\bibnamefont
  {Haferkamp}}, \bibinfo {author} {\bibfnamefont {D.}~\bibnamefont
  {Hangleiter}}, \bibinfo {author} {\bibfnamefont {J.}~\bibnamefont {Eisert}},\
  and\ \bibinfo {author} {\bibfnamefont {M.}~\bibnamefont {Gluza}},\ }\bibfield
   {title} {\bibinfo {title} {Contracting projected entangled pair states is
  average-case hard},\ }\href
  {https://doi.org/10.1103/PhysRevResearch.2.013010} {\bibfield  {journal}
  {\bibinfo  {journal} {Physical Review Research}\ }\textbf {\bibinfo {volume}
  {2}},\ \bibinfo {pages} {013010} (\bibinfo {year} {2020})},\ \bibinfo {note}
  {publisher: American Physical Society}\BibitemShut {NoStop}%
\bibitem [{\citenamefont {Deng}\ \emph
  {et~al.}(2017{\natexlab{b}})\citenamefont {Deng}, \citenamefont {Li},\ and\
  \citenamefont {Das~Sarma}}]{deng2017quantum}%
  \BibitemOpen
  \bibfield  {author} {\bibinfo {author} {\bibfnamefont {D.-L.}\ \bibnamefont
  {Deng}}, \bibinfo {author} {\bibfnamefont {X.}~\bibnamefont {Li}},\ and\
  \bibinfo {author} {\bibfnamefont {S.}~\bibnamefont {Das~Sarma}},\ }\bibfield
  {title} {\bibinfo {title} {Quantum entanglement in neural network states},\
  }\href {https://doi.org/10.1103/PhysRevX.7.021021} {\bibfield  {journal}
  {\bibinfo  {journal} {Phys. Rev. X}\ }\textbf {\bibinfo {volume} {7}},\
  \bibinfo {pages} {021021} (\bibinfo {year} {2017}{\natexlab{b}})}\BibitemShut
  {NoStop}%
\bibitem [{\citenamefont {Chen}\ \emph {et~al.}(2018)\citenamefont {Chen},
  \citenamefont {Cheng}, \citenamefont {Xie}, \citenamefont {Wang},\ and\
  \citenamefont {Xiang}}]{PhysRevB.97.085104}%
  \BibitemOpen
  \bibfield  {author} {\bibinfo {author} {\bibfnamefont {J.}~\bibnamefont
  {Chen}}, \bibinfo {author} {\bibfnamefont {S.}~\bibnamefont {Cheng}},
  \bibinfo {author} {\bibfnamefont {H.}~\bibnamefont {Xie}}, \bibinfo {author}
  {\bibfnamefont {L.}~\bibnamefont {Wang}},\ and\ \bibinfo {author}
  {\bibfnamefont {T.}~\bibnamefont {Xiang}},\ }\bibfield  {title} {\bibinfo
  {title} {Equivalence of restricted boltzmann machines and tensor network
  states},\ }\href {https://doi.org/10.1103/PhysRevB.97.085104} {\bibfield
  {journal} {\bibinfo  {journal} {Phys. Rev. B}\ }\textbf {\bibinfo {volume}
  {97}},\ \bibinfo {pages} {085104} (\bibinfo {year} {2018})}\BibitemShut
  {NoStop}%
\bibitem [{\citenamefont {Levine}\ \emph {et~al.}(2019)\citenamefont {Levine},
  \citenamefont {Sharir}, \citenamefont {Cohen},\ and\ \citenamefont
  {Shashua}}]{levine2019quantum}%
  \BibitemOpen
  \bibfield  {author} {\bibinfo {author} {\bibfnamefont {Y.}~\bibnamefont
  {Levine}}, \bibinfo {author} {\bibfnamefont {O.}~\bibnamefont {Sharir}},
  \bibinfo {author} {\bibfnamefont {N.}~\bibnamefont {Cohen}},\ and\ \bibinfo
  {author} {\bibfnamefont {A.}~\bibnamefont {Shashua}},\ }\bibfield  {title}
  {\bibinfo {title} {Quantum entanglement in deep learning architectures},\
  }\href {https://doi.org/10.1103/PhysRevLett.122.065301} {\bibfield  {journal}
  {\bibinfo  {journal} {Phys. Rev. Lett.}\ }\textbf {\bibinfo {volume} {122}},\
  \bibinfo {pages} {065301} (\bibinfo {year} {2019})}\BibitemShut {NoStop}%
\bibitem [{\citenamefont {Pastori}\ \emph {et~al.}(2019)\citenamefont
  {Pastori}, \citenamefont {Kaubruegger},\ and\ \citenamefont
  {Budich}}]{pastori_generalized_2019}%
  \BibitemOpen
  \bibfield  {author} {\bibinfo {author} {\bibfnamefont {L.}~\bibnamefont
  {Pastori}}, \bibinfo {author} {\bibfnamefont {R.}~\bibnamefont
  {Kaubruegger}},\ and\ \bibinfo {author} {\bibfnamefont {J.~C.}\ \bibnamefont
  {Budich}},\ }\bibfield  {title} {\bibinfo {title} {Generalized transfer
  matrix states from artificial neural networks},\ }\href
  {https://doi.org/10.1103/PhysRevB.99.165123} {\bibfield  {journal} {\bibinfo
  {journal} {Physical Review B}\ }\textbf {\bibinfo {volume} {99}},\ \bibinfo
  {pages} {165123} (\bibinfo {year} {2019})},\ \bibinfo {note} {publisher:
  American Physical Society}\BibitemShut {NoStop}%
\bibitem [{\citenamefont {Borin}\ and\ \citenamefont
  {Abanin}(2020)}]{borin_approximating_2020}%
  \BibitemOpen
  \bibfield  {author} {\bibinfo {author} {\bibfnamefont {A.}~\bibnamefont
  {Borin}}\ and\ \bibinfo {author} {\bibfnamefont {D.~A.}\ \bibnamefont
  {Abanin}},\ }\bibfield  {title} {\bibinfo {title} {Approximating power of
  machine-learning ansatz for quantum many-body states},\ }\href
  {https://doi.org/10.1103/PhysRevB.101.195141} {\bibfield  {journal} {\bibinfo
   {journal} {Physical Review B}\ }\textbf {\bibinfo {volume} {101}},\ \bibinfo
  {pages} {195141} (\bibinfo {year} {2020})},\ \bibinfo {note} {publisher:
  American Physical Society}\BibitemShut {NoStop}%
\bibitem [{\citenamefont {Park}\ and\ \citenamefont
  {Kastoryano}(2020)}]{park_are_2020}%
  \BibitemOpen
  \bibfield  {author} {\bibinfo {author} {\bibfnamefont {C.-Y.}\ \bibnamefont
  {Park}}\ and\ \bibinfo {author} {\bibfnamefont {M.~J.}\ \bibnamefont
  {Kastoryano}},\ }\bibfield  {title} {\bibinfo {title} {Are neural quantum
  states good at solving non-stoquastic spin {Hamiltonians}?},\ }\href
  {http://arxiv.org/abs/2012.08889} {\bibfield  {journal} {\bibinfo  {journal}
  {arXiv:2012.08889 [cond-mat, physics:quant-ph]}\ } (\bibinfo {year}
  {2020})},\ \bibinfo {note} {arXiv: 2012.08889}\BibitemShut {NoStop}%
\bibitem [{Note1()}]{Note1}%
  \BibitemOpen
  \bibinfo {note} {This is the classical definition of a neural network.
  However, some of the models used today slightly deviate from it, e.g.,
  self-attention modules use bilinear operations in addition to affine ones.
  While our proofs for the main results only consider the classical definition,
  extending it to support many of these variants is trivial.}\BibitemShut
  {Stop}%
\bibitem [{\citenamefont {Nair}\ and\ \citenamefont {Hinton}(2010)}]{ReLU}%
  \BibitemOpen
  \bibfield  {author} {\bibinfo {author} {\bibfnamefont {V.}~\bibnamefont
  {Nair}}\ and\ \bibinfo {author} {\bibfnamefont {G.~E.}\ \bibnamefont
  {Hinton}},\ }\bibfield  {title} {\bibinfo {title} {Rectified linear units
  improve restricted boltzmann machines},\ }in\ \href
  {https://icml.cc/Conferences/2010/papers/432.pdf} {\emph {\bibinfo
  {booktitle} {ICML}}}\ (\bibinfo {year} {2010})\ pp.\ \bibinfo {pages}
  {807--814}\BibitemShut {NoStop}%
\bibitem [{\citenamefont {Dugas}\ \emph {et~al.}(2000)\citenamefont {Dugas},
  \citenamefont {Bengio}, \citenamefont {Bélisle}, \citenamefont {Nadeau},\
  and\ \citenamefont {Garcia}}]{Softplus}%
  \BibitemOpen
  \bibfield  {author} {\bibinfo {author} {\bibfnamefont {C.}~\bibnamefont
  {Dugas}}, \bibinfo {author} {\bibfnamefont {Y.}~\bibnamefont {Bengio}},
  \bibinfo {author} {\bibfnamefont {F.}~\bibnamefont {Bélisle}}, \bibinfo
  {author} {\bibfnamefont {C.}~\bibnamefont {Nadeau}},\ and\ \bibinfo {author}
  {\bibfnamefont {R.}~\bibnamefont {Garcia}},\ }\bibfield  {title} {\bibinfo
  {title} {Incorporating second-order functional knowledge for better option
  pricing},\ }in\ \href
  {http://papers.nips.cc/paper/1920-incorporating-second-order-functional-knowledge-for-better-option-pricing}
  {\emph {\bibinfo {booktitle} {NIPS}}}\ (\bibinfo {year} {2000})\ pp.\
  \bibinfo {pages} {472--478}\BibitemShut {NoStop}%
\bibitem [{\citenamefont {Cohen}\ \emph {et~al.}(2018)\citenamefont {Cohen},
  \citenamefont {Sharir}, \citenamefont {Levine}, \citenamefont {Tamari},
  \citenamefont {Yakira},\ and\ \citenamefont {Shashua}}]{cohen2018analysis}%
  \BibitemOpen
  \bibfield  {author} {\bibinfo {author} {\bibfnamefont {N.}~\bibnamefont
  {Cohen}}, \bibinfo {author} {\bibfnamefont {O.}~\bibnamefont {Sharir}},
  \bibinfo {author} {\bibfnamefont {Y.}~\bibnamefont {Levine}}, \bibinfo
  {author} {\bibfnamefont {R.}~\bibnamefont {Tamari}}, \bibinfo {author}
  {\bibfnamefont {D.}~\bibnamefont {Yakira}},\ and\ \bibinfo {author}
  {\bibfnamefont {A.}~\bibnamefont {Shashua}},\ }\href@noop {} {\bibinfo
  {title} {Analysis and design of convolutional networks via hierarchical
  tensor decompositions}} (\bibinfo {year} {2018}),\ \Eprint
  {https://arxiv.org/abs/1705.02302} {arXiv:1705.02302 [cs.LG]} \BibitemShut
  {NoStop}%
\bibitem [{\citenamefont {Mhaskar}\ \emph {et~al.}(2017)\citenamefont
  {Mhaskar}, \citenamefont {Liao},\ and\ \citenamefont {Poggio}}]{Poggio2017}%
  \BibitemOpen
  \bibfield  {author} {\bibinfo {author} {\bibfnamefont {H.}~\bibnamefont
  {Mhaskar}}, \bibinfo {author} {\bibfnamefont {Q.}~\bibnamefont {Liao}},\ and\
  \bibinfo {author} {\bibfnamefont {T.}~\bibnamefont {Poggio}},\ }\bibfield
  {title} {\bibinfo {title} {When and why are deep networks better than shallow
  ones?},\ }\href {https://ojs.aaai.org/index.php/AAAI/article/view/10913}
  {\bibfield  {journal} {\bibinfo  {journal} {Proceedings of the AAAI
  Conference on Artificial Intelligence}\ }\textbf {\bibinfo {volume} {31}}
  (\bibinfo {year} {2017})}\BibitemShut {NoStop}%
\bibitem [{\citenamefont {Yarotsky}(2017)}]{Yarotsky2017}%
  \BibitemOpen
  \bibfield  {author} {\bibinfo {author} {\bibfnamefont {D.}~\bibnamefont
  {Yarotsky}},\ }\bibfield  {title} {\bibinfo {title} {Error bounds for
  approximations with deep relu networks},\ }\href
  {https://doi.org/https://doi.org/10.1016/j.neunet.2017.07.002} {\bibfield
  {journal} {\bibinfo  {journal} {Neural Networks}\ }\textbf {\bibinfo {volume}
  {94}},\ \bibinfo {pages} {103 } (\bibinfo {year} {2017})}\BibitemShut
  {NoStop}%
\bibitem [{\citenamefont {Telgarsky}(2017)}]{telgarsky17a}%
  \BibitemOpen
  \bibfield  {author} {\bibinfo {author} {\bibfnamefont {M.}~\bibnamefont
  {Telgarsky}},\ }\bibfield  {title} {\bibinfo {title} {Neural networks and
  rational functions},\ }in\ \href
  {http://proceedings.mlr.press/v70/telgarsky17a.html} {\emph {\bibinfo
  {booktitle} {Proceedings of the 34th International Conference on Machine
  Learning}}},\ \bibinfo {series} {Proceedings of Machine Learning Research},
  Vol.~\bibinfo {volume} {70},\ \bibinfo {editor} {edited by\ \bibinfo {editor}
  {\bibfnamefont {D.}~\bibnamefont {Precup}}\ and\ \bibinfo {editor}
  {\bibfnamefont {Y.~W.}\ \bibnamefont {Teh}}}\ (\bibinfo  {publisher} {PMLR},\
  \bibinfo {address} {International Convention Centre, Sydney, Australia},\
  \bibinfo {year} {2017})\ pp.\ \bibinfo {pages} {3387--3393}\BibitemShut
  {NoStop}%
\bibitem [{Note2()}]{Note2}%
  \BibitemOpen
  \bibinfo {note} {The bounds on the two contraction schemes serve to highlight
  different characteristics. By using the sequential scheme, it is demonstrated
  that NN can approximate MPS with the same optimal runtime when not accounting
  for parallelization. By employing the parallel scheme, it demonstrates that a
  logarithmic depth is sufficient for approximating MPS, while also better
  utilizing parallel execution as supported by modern GPUs.}\BibitemShut
  {Stop}%
\bibitem [{\citenamefont {Hastings}(2007)}]{hastings_area_2007}%
  \BibitemOpen
  \bibfield  {author} {\bibinfo {author} {\bibfnamefont {M.~B.}\ \bibnamefont
  {Hastings}},\ }\bibfield  {title} {\bibinfo {title} {An area law for
  one-dimensional quantum systems},\ }\href
  {https://doi.org/10.1088/1742-5468/2007/08/P08024} {\bibfield  {journal}
  {\bibinfo  {journal} {Journal of Statistical Mechanics: Theory and
  Experiment}\ }\textbf {\bibinfo {volume} {2007}},\ \bibinfo {pages} {P08024}
  (\bibinfo {year} {2007})},\ \bibinfo {note} {publisher: IOP
  Publishing}\BibitemShut {NoStop}%
\bibitem [{\citenamefont {Arad}\ \emph {et~al.}(2013)\citenamefont {Arad},
  \citenamefont {Kitaev}, \citenamefont {Landau},\ and\ \citenamefont
  {Vazirani}}]{arad_area_2013}%
  \BibitemOpen
  \bibfield  {author} {\bibinfo {author} {\bibfnamefont {I.}~\bibnamefont
  {Arad}}, \bibinfo {author} {\bibfnamefont {A.}~\bibnamefont {Kitaev}},
  \bibinfo {author} {\bibfnamefont {Z.}~\bibnamefont {Landau}},\ and\ \bibinfo
  {author} {\bibfnamefont {U.}~\bibnamefont {Vazirani}},\ }\bibfield  {title}
  {\bibinfo {title} {An area law and sub-exponential algorithm for {1D}
  systems},\ }\href {http://arxiv.org/abs/1301.1162} {\bibfield  {journal}
  {\bibinfo  {journal} {arXiv:1301.1162 [cond-mat, physics:quant-ph]}\ }
  (\bibinfo {year} {2013})},\ \bibinfo {note} {arXiv: 1301.1162}\BibitemShut
  {NoStop}%
\bibitem [{\citenamefont {Schuch}\ and\ \citenamefont
  {Verstraete}(2017)}]{schuch_matrix_2017}%
  \BibitemOpen
  \bibfield  {author} {\bibinfo {author} {\bibfnamefont {N.}~\bibnamefont
  {Schuch}}\ and\ \bibinfo {author} {\bibfnamefont {F.}~\bibnamefont
  {Verstraete}},\ }\bibfield  {title} {\bibinfo {title} {Matrix product state
  approximations for infinite systems},\ }\href
  {http://arxiv.org/abs/1711.06559} {\bibfield  {journal} {\bibinfo  {journal}
  {arXiv:1711.06559 [cond-mat, physics:quant-ph]}\ } (\bibinfo {year}
  {2017})},\ \bibinfo {note} {arXiv: 1711.06559}\BibitemShut {NoStop}%
\bibitem [{\citenamefont {Dalzell}\ and\ \citenamefont
  {Brandão}(2019)}]{dalzell_locally_2019}%
  \BibitemOpen
  \bibfield  {author} {\bibinfo {author} {\bibfnamefont {A.~M.}\ \bibnamefont
  {Dalzell}}\ and\ \bibinfo {author} {\bibfnamefont {F.~G. S.~L.}\ \bibnamefont
  {Brandão}},\ }\bibfield  {title} {\bibinfo {title} {Locally accurate {MPS}
  approximations for ground states of one-dimensional gapped local
  {Hamiltonians}},\ }\href {https://doi.org/10.22331/q-2019-09-23-187}
  {\bibfield  {journal} {\bibinfo  {journal} {Quantum}\ }\textbf {\bibinfo
  {volume} {3}},\ \bibinfo {pages} {187} (\bibinfo {year} {2019})},\ \bibinfo
  {note} {publisher: Verein zur Förderung des Open Access Publizierens in den
  Quantenwissenschaften}\BibitemShut {NoStop}%
\bibitem [{\citenamefont {Sharir}\ and\ \citenamefont
  {Shashua}(2018)}]{sharir2018expressive}%
  \BibitemOpen
  \bibfield  {author} {\bibinfo {author} {\bibfnamefont {O.}~\bibnamefont
  {Sharir}}\ and\ \bibinfo {author} {\bibfnamefont {A.}~\bibnamefont
  {Shashua}},\ }\bibfield  {title} {\bibinfo {title} {On the expressive power
  of overlapping architectures of deep learning},\ }in\ \href@noop {} {\emph
  {\bibinfo {booktitle} {6th International Conference on Learning
  Representations (ICLR)}}}\ (\bibinfo {year} {2018})\BibitemShut {NoStop}%
\bibitem [{\citenamefont {Evenbly}\ and\ \citenamefont
  {Vidal}(2014{\natexlab{a}})}]{PhysRevLett.112.240502}%
  \BibitemOpen
  \bibfield  {author} {\bibinfo {author} {\bibfnamefont {G.}~\bibnamefont
  {Evenbly}}\ and\ \bibinfo {author} {\bibfnamefont {G.}~\bibnamefont
  {Vidal}},\ }\bibfield  {title} {\bibinfo {title} {Class of highly entangled
  many-body states that can be efficiently simulated},\ }\href
  {https://doi.org/10.1103/PhysRevLett.112.240502} {\bibfield  {journal}
  {\bibinfo  {journal} {Phys. Rev. Lett.}\ }\textbf {\bibinfo {volume} {112}},\
  \bibinfo {pages} {240502} (\bibinfo {year} {2014}{\natexlab{a}})}\BibitemShut
  {NoStop}%
\bibitem [{\citenamefont {Evenbly}\ and\ \citenamefont
  {Vidal}(2014{\natexlab{b}})}]{evenbly2014scaling}%
  \BibitemOpen
  \bibfield  {author} {\bibinfo {author} {\bibfnamefont {G.}~\bibnamefont
  {Evenbly}}\ and\ \bibinfo {author} {\bibfnamefont {G.}~\bibnamefont
  {Vidal}},\ }\bibfield  {title} {\bibinfo {title} {Scaling of entanglement
  entropy in the (branching) multiscale entanglement renormalization ansatz},\
  }\href@noop {} {\bibfield  {journal} {\bibinfo  {journal} {Physical Review
  B}\ }\textbf {\bibinfo {volume} {89}},\ \bibinfo {pages} {235113} (\bibinfo
  {year} {2014}{\natexlab{b}})}\BibitemShut {NoStop}%
\bibitem [{\citenamefont {Andrecut}(2009)}]{SVDImplementation}%
  \BibitemOpen
  \bibfield  {author} {\bibinfo {author} {\bibfnamefont {M.}~\bibnamefont
  {Andrecut}},\ }\bibfield  {title} {\bibinfo {title} {Parallel gpu
  implementation of iterative pca algorithms},\ }\href
  {https://doi.org/10.1089/cmb.2008.0221} {\bibfield  {journal} {\bibinfo
  {journal} {Journal of Computational Biology}\ }\textbf {\bibinfo {volume}
  {16}},\ \bibinfo {pages} {1593} (\bibinfo {year} {2009})},\ \bibinfo {note}
  {pMID: 19772385},\ \Eprint
  {https://arxiv.org/abs/https://doi.org/10.1089/cmb.2008.0221}
  {https://doi.org/10.1089/cmb.2008.0221} \BibitemShut {NoStop}%
\end{thebibliography}%


\ifdefined\ARXIV
\appendix
\section{Introduction to Tensor Networks}\label{app:tn_intro}

\begin{figure*}[h!]
    \centering
    \includegraphics[width=\linewidth,keepaspectratio]{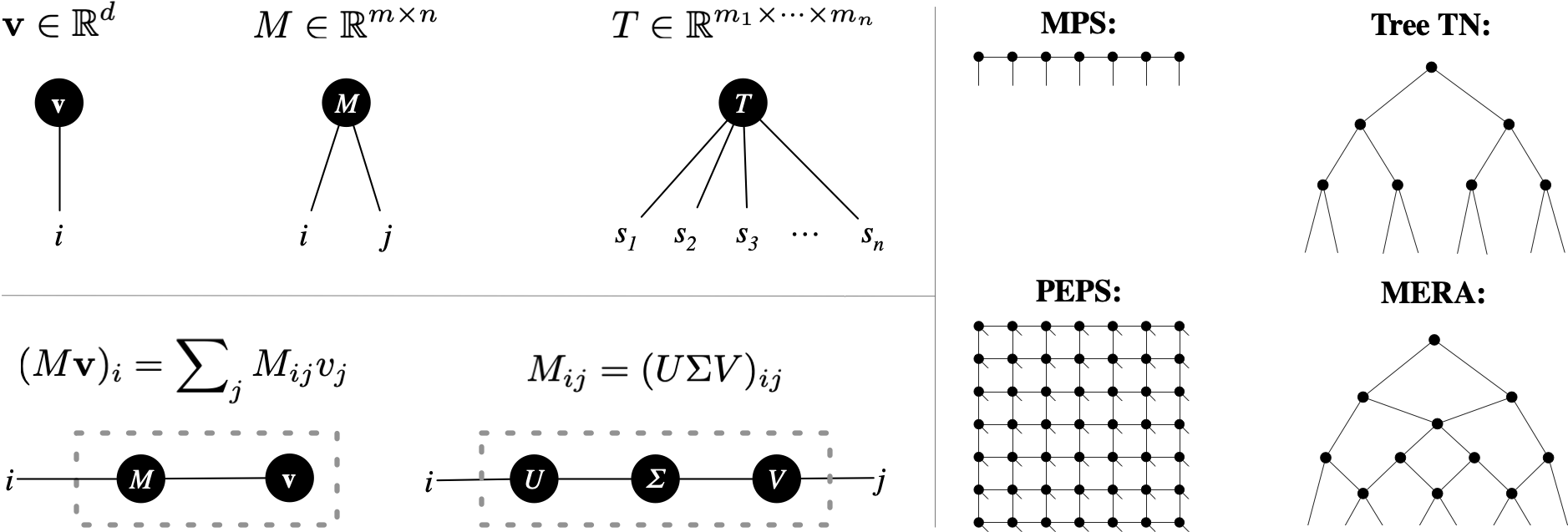}
    \caption{\label{fig:tn_intro}Illustrations of basic elements and operations of tensor networks, as well as common tensor networks types.}
\end{figure*}

Here we give a brief introduction to the basic concepts of tensor networks~(TN). See Fig.~\ref{fig:tn_intro} for the accompanying illustrations. TN is a graphical notation for describing common tensor operations and factorization schemes. Nodes in the graph represent tensors, where edges correspond to indices, ranging from vectors (top left) and matrices (top middle) to arbitrary high-dimensional tensors (top right). Connected nodes represent tensor contractions, i.e., a summation over matching indices of the products of all tensor nodes in the graph, e.g., matrix-vector multiplication (bottom left). Tensor networks are useful for describing tensor factorizations, e.g., SVD factorization of matrices (bottom right). The most commonly used forms of TN are Matrix Product States~(MPS), Tree Tensor Networks~(TTN), Projected Entangled Pair States~(PEPS), and Multi-scale Entanglement Renormalization Ansatz~(MERA).

\begin{figure*}[h]
    \centering
    \includegraphics[width=\textwidth,keepaspectratio]{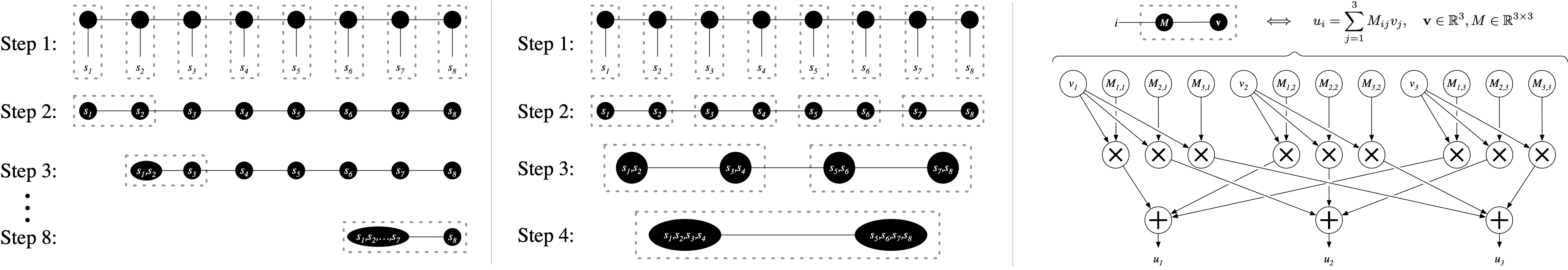}
    \caption{\label{fig:mps_contraction}
        \textbf{(left)} Sequential contraction scheme for Matrix Product States: At step 1, we map indices $d_1, \ldots, d_8$ to their corresponding matrices (or vectors at boundaries), a $O(d \chi^2)$-time operation. In each of the following steps, we contract a boundary vector with its neighboring matrix node, a $O(\chi^2)$-time operation, amounting to a total of $O(N d \chi^2)$ for the entire contraction, performed in $N$ steps. 
        \textbf{(middle)}~Parallel\protect\footnotemark contraction scheme for Matrix Product States: Following step 1 as in the sequential contraction, we contract pairs of neighboring nodes in parallel, each an $O(\chi^3)$-time operation, amounting to a total of $O(N (d + \chi) \chi^2)$ for the entire contraction, performed in $\log_2(N)$ steps.
        \textbf{(right)}~Illustration of a simple contraction scheme, in this case matrix-vector multiplication, as an arithmetic circuit.}
\end{figure*}

\footnotetext{When parallelizing across sites, the \emph{effective} run-time in practice depends mostly on the number of steps, i.e., $\log_2 N$, and $\chi^2$ rather than $\chi^3$ because each matrix multiplication itself can be parallelized across the coordinates of the output matrix, resulting in $O(\chi^2 \log N)$.}

The complexity of contracting a TN exactly is dependent on its contraction order. While finding the optimal contraction order for an arbitrary TN is known to be NP-complete, for many common TN forms, e.g. Matrix Product States, efficient algorithms exist. Two such contraction schemes are the sequential and parallel contractions that are depicted in Fig.~\ref{fig:mps_contraction}.

\section{Related Works on Approximating Polynomial Functions with Neural Networks}\label{app:related_theoretical_results}

When examining the ability of NN to approximate polynomials, one can notice that while the weighted sum operation is straitforward for NN, the product operation is not trivially simulated by NN and has been the topic of several works \citep{Poggio2017,Yarotsky2017,telgarsky17a} in the context of the approximation power of NN.
Nevertheless, we could not base our approximation scheme on these claims without attaining worse bounds.
The most recent result~\citep{Yarotsky2017} on approximating products with NN demonstrates a construction with a width and a depth at most $O(\log(\nicefrac{M}{\epsilon})) $ such that $\max_{x,y\in[-M,M]}\abs{\mathrm{NN}(x,y) - x \cdot y} < \epsilon$.
While this impressive rate of approximation is sufficient for many purposes, it is less suitable for quantum states representation.

Consider for example an arbitrary $N$-qubit system, then due to normalization at least half of its wave-function amplitudes are, in modulus, less than $2^{-N/2}$, which entails $\epsilon < 2^{-N/2}$ for a meaningful approximation.
Thus, using this construction would require at least $\mathrm{poly}(N)$ width and depth for every product operation, resulting in a multiplicative polynomial penalty to the runtime.
In practice, this polynomial penalty would have major ramifications. To put this in perspective, a $10{\times}10$ two-dimensional system would require at least hundreds of NN layers regardless of the complexity of the TNS.

\section{Proofs of Claims on the Relationship Between the Fidelity and the Infinity Norm}\label{app:fidelity}
\subsection{Proof of Claim~\ref{claim:good_fidelity}}
Let $\psi$ and $\phi$ be two WF with non-zero magnitudes everywhere, such that their logarithm is well defined, i.e., $\ln \psi(s) = a(s) + i \cdot b(s)$ and $\ln \phi(s) = c(s) + i \cdot d(s)$. Assume $\norm{\ln \psi - \ln \phi}_\infty < \epsilon < 1$.

We begin by finding a lower bound for the inner product:
\begin{align*}
	\abs{\langle \psi | \phi \rangle} &\geq \abs{Re(\langle \psi | \phi \rangle)} = \abs{\sum_s e^{a(s) + c(s)} \cos(d(s) - b(s))} \\
	&= \abs{\sum_s \abs{\psi(s)}^2 e^{c(s) - a(s)} \cos(d(s) - b(s))} \\
	&\geq \abs{\sum_s \abs{\psi(s)}^2 e^{-\epsilon} \cos(\epsilon)} = \norm{\psi}_2^2 e^{-\epsilon} \cos(\epsilon)
\end{align*}
and similarly we can find an upper bound for the norm of $\phi$:
\begin{align*}
	\norm{\phi}_2^2 &= \sum_s \abs{\phi(s)}^2 = \sum_s \abs{\psi(s)}^2 e^{2(c(s) - a(s))}\leq \norm{\psi}_2^2 e^{2\epsilon}
\end{align*}

Using the above, we can find a lower bound for the fidelity:
\begin{align*}
	F(\psi,\phi) &= \frac{\abs{\langle \psi | \phi \rangle}^2}{\norm{\psi}_2^2 \norm{\phi}_2^2} \geq (\cos \epsilon)^2 \geq 1 - \epsilon^2
\end{align*}

\subsection{Proof of Claim~\ref{claim:bad_fidelity}}
Let $s' \in 2^N$ an arbitrary point and define:
\begin{align*}
	\psi(s) &= \begin{cases}
		1 & s = s' \\
		0 & otherwise
	\end{cases} \\
	\phi(s) &= \begin{cases}
		1 - \epsilon & s = s' \\
		\epsilon & otherwise
	\end{cases}
\end{align*}
Clearly $\norm{\psi - \phi}_\infty \leq \epsilon$, and yet:
\begin{align*}
	F(\psi, \phi) &= \frac{1 - 2\epsilon + \epsilon^2}{1 - 2\epsilon + 2^N \epsilon^2} \\
	&\approx \begin{cases}
		2^{-N} & \epsilon^{-2} \ll 2^N \\
		1 - 2^N \epsilon^2 & \epsilon^{-2} \gg 2^N
	\end{cases}
\end{align*}

\section{Proof of Theorem \ref{th:main}}\label{app:proof}

In this section we describe the proof of Theorem \ref{th:main}. We begin by providing a sketch of the proof, followed by the full proof.
As mentioned in the main text, we prove the theorem in two steps. First, prove the theorem for the case of non-negative AC. Second, reduce the general complex case to the non-negative case.

\subsection{Proof Sketch}
The proof is based on two steps. First, we show that AC with non-negative parameters and inputs can be exactly reconstructed with NN with real parameters and softplus activation functions. Let $o_1 = \log(x_1), o_2 = \log(x_2)$ for $x_1,x_2 \geq 0$. Then, working in log-space, multiplication becomes summation, i.e., $\log(x_1 \cdot x_2) = o_1 + o_2$, making input-input multiplication trivial for NN, unlike before. For every input-parameter multiplication, i.e., a sum-node edge in the AC graph, we add an auxiliary neuron with a single input. The AC's parameters are stored in the bias terms of these auxiliary neurons, adding $m$ nodes to the NN but with negligible effect on runtime (number of edges). For summation, softplus activations arise naturally:
\begin{align}
\log(x_1 + x_2) &= \log(\exp(o_1) + \exp(o_2)) \nonumber\\
&= o_1 + \log\left(1 + \exp(o_2 - o_1) \right) \nonumber\\
&= o_1 + \softplus(o_2 - o_1). \label{eq:log_sum_exp_to_softplus}
\end{align}
For log-space summation of $n$ inputs, we can decompose it as a binary tree, which gives the $\log(m)$ correction to the depth of the network. With both log-space NN analogs in place, a non-negative AC can be exactly reproduce with same asymptotic time complexity.

For the second step, we reduce the general complex case to the non-negative case. A real number $x\in\R$ can be represented with a redundant representation of two non-negative numbers $x_{\mathrm{+}},x_{\mathrm{-}}\geq 0$ by $x = x_{\mathrm{+}} - x_{\mathrm{-}}$. Addition and multiplication can be applied directly on this representation:
\begin{align*}
x + y &= (x_{\mathrm{+}} + y_{\mathrm{+}}) - (x_{\mathrm{-}} + y_{\mathrm{-}}) \\
x \cdot y &= (x_{\mathrm{+}} \cdot y_{\mathrm{+}} + x_{\mathrm{-}} \cdot y_{\mathrm{-}}) - (x_{\mathrm{-}} \cdot y_{\mathrm{+}} + x_{\mathrm{+}} \cdot y_{\mathrm{-}})
\end{align*}
Thus, a real AC can be expressed as the difference of two non-negative AC, and a complex AC by representing the real and imaginary parts in this fashion. Finally, to compute the logarithm of this redundant complex representation, i.e., the log-magnitude and phase, we employ various univariate approximation schemes. Since these two operations are smooth and used only at the end of the network, it results in the additive term $c(\epsilon,m,W_{\max},f_{\min})$, which is merely logarithmic in the number of edges of the AC, and double logarithmic with respect to the magnitudes of the weights and the WF amplitudes. Due to these weak dependencies of the target AC, it allows for an approximation with a practically arbitrary precision.

\subsection{Non-negative Case}
For the first step, we assume an AC with non-negative inputs and parameters. The inputs and AC parameters are transformed to their log-value, where we extend the real-line with $\pm \infty$ and represent $\log(0) = -\infty$. For most practical considerations, $-\infty$ could be substituted with a large but finite negative constant.

In our NN construction, we freely use the identity instead of a softplus activation function when it is more convenient. We can do so because the identity operation can be simulated with arbitrary precision using the weighted sum of just two neurons with softplus activations:
\begin{align*}
    x &= \max(x, 0) - \max(-x, 0),\\
    \max(x, 0) &= \lim_{\delta\to\infty} \frac{\softplus(\delta x)}{\delta} =\lim_{\delta\to\infty} \frac{1}{\delta} \ln(1+\exp(\delta x)),\\
    &= \lim_{\delta\to\infty} \begin{cases} \frac{1}{\delta} \overbrace{\ln(1+\exp(\delta x))}^{\to 0} & x \leq 0 \\
     x + \frac{1}{\delta} \overbrace{\ln(1+\exp(-\delta x))}^{\to 0} & x > 0
    \end{cases}, \\
    \Rightarrow x &= \lim_{\delta\to\infty} \frac{\softplus(\delta x) - \softplus(-\delta x)}{\delta}.
\end{align*}
The above workaround can at most double the number of neurons and edges in our construction, and thus does not affect our asymptotic bounds.

Every product node with $k$ in-edges in the AC is replaced by a neuron with $k$ in-edges, whose weights are set to $1$ and bias to $0$, representing multiplication in log-space, i.e., $\log(\prod_{i=1}^k x_i) = \sum_{i=1}^k o_i $, where $\{o_i = \exp(x_i)\}_{i=1}^k$ are the log-values of the connected nodes.

Every weighted-sum node with $k$ in-edges and parameterized by $\w \in \R^k_{\geq 0}$ is replaced by the following NN sub-graph of $O(k)$ nodes and $O(k)$ edges. Every input-parameter multiplication term, i.e., $w_i \cdot x_i$, is represented by a single neuron with a single in-edge with weights set to $0$ and bias set to $w_i$, resulting in $p_i \equiv \log(w_i \cdot x_i) = w_i + o_i$. Without loosing our generality, assume $k = 2^t$ for some $t \in \N$, and so we can decompose $\sum_{i=1}^k p_i$ as a complete binary tree of depth $t$, $2k - 1$ nodes, and $2k - 1$ in-edges in total. Each node in the tree represent a binary addition, which can be realized with 2 neurons, one with softplus activation and one with identity:
\begin{align*}
\log(x_1 + x_2) &= \log(\exp(o_1) + \exp(o_2)) \\
&= o_1 + \log\left(1 + \exp(o_2 - o_1) \right) \\
&= o_1 + \softplus(o_2 - o_1).
\end{align*}

Applying the above transformations to a non-negative AC with $n$ nodes, $m$ edges, and depth $l$ results in a NN of depth $l \log(m)$ with $O(n + m)$ nodes and $O(m)$ edges, concluding the proof of the first step.

\subsection{Complex Case}

For the second step, we begin initially by transforming a complex AC into four distinct non-negative AC graphs, representing the following four ``parts'' of a complex number: positive real, negative real, positive imaginary, and negative imaginary.

Every real number $x \in \R$ can be represented with the redundant form $x = x_+ - x_-$, where $x_+,x_- \in \R_{\geq 0}$. Multiplication and addition can be performed directly within that representation using the following identities:
\begin{align*}
    x + y &= [x_+ + y_+] - [x_- - y_-], \\
    x \cdot y &= [x_+ \cdot y_+ + x_- \cdot y_-] - [x_+ \cdot y_- + x_- \cdot y_+].
\end{align*}
Similarly, a complex number $z \in \C$ can be represented with four components, $z = z_{\mathrm{re},+} - z_{\mathrm{re},-}+i\cdot(z_{\mathrm{im},+} - z_{\mathrm{im},-})$, where $z_{\mathrm{re},+}, z_{\mathrm{re},-}, z_{\mathrm{im},+}, z_{\mathrm{im},-} \in \R_{\geq 0}$.

Given a complex AC with $m$ edges, $n$ nodes, and of depth $l$, we can use the above redundant representation for its inputs, parameters, and intermediate computations. Propagating the operations with the above identities through the complex AC graph, results in four non-negative AC, each with $O(m)$ edges, $O(n)$ nodes, and of depth $O(l)$, denoting each component of the complex AC's output, i.e., $AC(z) = AC(\hat z)_{\mathrm{re},+} -  AC(\hat z)_{\mathrm{re},-} + i \cdot \left( AC(\hat z)_{\mathrm{im},+} - AC(\hat z)_{\mathrm{im},-}\right)$, where $\hat{z} = (z_{\mathrm{re},+}, z_{\mathrm{re},-}, z_{\mathrm{im},+}, z_{\mathrm{im},-})$. The logarithm of each of these non-negative AC can be represented with a NN according to the first step.

What remains is to convert the redundant representation to a log-polar form, i.e., ${\log(z) = \log(\abs{z}) + i\cdot \arg(z)}$, per the desired output described in Theorem~\ref{th:main}. We employ various approximation techniques to simulate this operation. In the following we denote the components of the redundant representation and its log-value by $o_{\mathrm{re},+} = \ln z_{\mathrm{re},+}$, $o_{\mathrm{re},-} = \ln z_{\mathrm{re},-}$, $o_{\mathrm{im},+} = \ln z_{\mathrm{im},+}$, and $o_{\mathrm{im},-} = \ln z_{\mathrm{im},-}$.

\subsubsection{Estimating $\log\abs{z}$}

In this sub-section, we describe the estimation of $\log(\abs{z})$ by softplus networks.

$\log(\abs{z})$ can be expressed with respect to the redundant representation's components as:
\begin{align}
   \log\abs{z} &= \ln\left(\sqrt{\abs{z_{\mathrm{re}}}^2 + \abs{z_{\mathrm{im}}}^2} \right),\nonumber\\
   &= \ln \abs{z_{\mathrm{re}}} + \frac{1}{2} \ln\left(1 + \exp\left(2\ln\abs{z_{\mathrm{im}}} -  2\ln \abs{z_{\mathrm{re}}} \right)\right), \nonumber\\
   &= \ln \abs{z_{\mathrm{re}}} + \frac{1}{2} \softplus(2\ln\abs{z_{\mathrm{im}}} -  2\ln \abs{z_{\mathrm{re}}}), \label{eq:log_z}
\end{align}
where $z_{\mathrm{re}} = z_{\mathrm{re},+} - z_{\mathrm{re},-}$  and $z_{\mathrm{im}} = z_{\mathrm{im},+} - z_{\mathrm{im},-}$.

In the rest of this sub-section we focus on the approximation of $\ln\abs{z_{\mathrm{re}}}$, where the same methods can be applied for $\ln\abs{z_{\mathrm{im}}}$.
We begin by defining $o_{\mathrm{re},\max} = \max(o_{\mathrm{re},+}, o_{\mathrm{re},-})$ and $o_{\mathrm{re},\min} = \min(o_{\mathrm{re},+}, o_{\mathrm{re},-})$, and similarly for the imaginary part. Recall that $\max(x,y) = y + \max(x-y,0)$ and $\min(x,y) = y - \max(y-x,0)$, and so both can be approximated to arbitrary precision with softplus networks. With that, we can write:
\begin{align*}
    \ln \abs{z_{\mathrm{re}}} &= \ln \left(\max(z_{\mathrm{re},+},z_{\mathrm{re}, -}) - \min(z_{\mathrm{re},+}, z_{\mathrm{re}, -})\right)\\
    &= \ln\left(\exp(o_{\mathrm{re},\max}) - \exp(o_{\mathrm{re},\min})\right),\\
    &= o_{\mathrm{re},\min} + \ln\left(\exp(o_{\mathrm{re},\max} - o_{\mathrm{re},\min}) - 1 \right),\\
    &= o_{\mathrm{re},\min} + \softplus^{-1}(o_{\mathrm{re},\max} - o_{\mathrm{re},\min}),
\end{align*}
where $\softplus^{-1}$ is the inverse of the softplus function. To approximate the inverse, we employ two strategies: (i)~for large values, $\softplus^{-1}(x) \approx x$ to a high precision, and (ii)~for smaller values, we estimate the inverse using root-finding algorithms, and specifically, the bisection method. 

Let $\epsilon > 0$, and $x = o_{\mathrm{re},\max} - o_{\mathrm{re},\min}$. For $x > x_{\mathrm{large}} \equiv - \ln(1 - \exp(-\epsilon))$ it holds that $\abs{x - \softplus^{-1}(x)} < \epsilon$. For realizing the bisection method, we first set the initial search range for $y^* = \softplus^{-1}(x)$. $y^*_{\max}$ can be set to $x_{\mathrm{large}}$ because $\softplus^{-1}(x) \leq x$. For $y^*_{\min}$ we can bound the minimal value of $x$ as follows
\begin{align*}
    x &= o_{\mathrm{re},\max} - o_{\mathrm{re},\min} = \ln \left(\frac{z_{\mathrm{re},\max}}{z_{\mathrm{re},\min}}\right) = \ln \left(\frac{\abs{z_{\mathrm{re}}} + z_{\mathrm{re},\min}}{z_{\mathrm{re},\min}}\right) \\
    &= \ln \left(\frac{\abs{z_{\mathrm{re}}}}{z_{\mathrm{re},\min}} + 1\right) \geq \ln \left(\frac{f_{\min}}{z_{\mathrm{re},\min}} + 1\right).
\end{align*}
Next, we upper bound the value of $z_{\mathrm{re},\min}$ by finding an upper bound on the value of a generic non-negative AC with $m$ edges. First, we replace every non-zero weight with the maximal weight in the graph. Then, we can replace every weighted sum with $v(\s) = \sum_{(u,v) \in E} W_{u,v}u(\s) \leq {\abs{\{(u,v) \in E\}}}\left(\max_{e \in E} W_e\right)\left(\max_{(u,v) \in E}  u(\s)\right)$. Finally, we can prove by induction along the topological order of the graph that the output of every sub-graph of $m'$ edges is upper bounded by $\left(m'\max_{(v,u) \in E} \abs{W_{v,u}}\right)^{m'}$. Thereby, we can set $x_{\min} \equiv \ln \left(\frac{f_{\min}}{\left(m W_{\max}\right)^m} + 1\right)$, and thus $y^*_{\min} \equiv \ln \left(\frac{f_{\min}}{\left(m W_{\max}\right)^m}\right)$.

To simulate the bisection algorithm, we define the approximate Heaviside function by $H_\delta(x) \equiv \max(\frac{x}{2\delta} + \frac{1}{2},0) - \max(\frac{x}{2\delta} - \frac{1}{2},0)$ that satisfies $H = \lim_{\delta \to 0} H_\delta$,  and use the following recursive update rule for $T \equiv \lceil\log_2 (\nicefrac{y^*_{\max} - y^*_{\min}}{\epsilon})\rceil$ steps:
\begin{align*}
    m_i &\equiv \frac{y_{i-1,min} + y_{i-1,max}}{2},\\
    c_i &\equiv H_\delta(\softplus(m) - x) \\
    y_{i,\min} &\equiv c_i y_{i-1,\min} + (1 - c_i) m_i, \\
    y_{i,\max} &\equiv c_i m_i + (1 - c_i) y_{i-1,\max},
\end{align*}
where the multiplications are approximated according to \citet{Yarotsky2017},  which requires an additional $O(\ln(\nicefrac{\max\{\abs{y^*_{\max}}, \abs{y^*_{\min}}\}}{\tilde\epsilon}))$ edges and depth per multiplication, where $\tilde\epsilon \equiv \nicefrac{\epsilon}{8T}$.
The usual bisection method relies on the exact Heaviside function, however, if   $\delta$ is chosen to be small enough, then it too satisfies the range halving property, i.e., it holds that $y_{i,max} - y_{i,min} = \frac{y_{i-1,max} - y_{i-1,min}}{2}$ and $\softplus^{-1}(x) \in [y_{i,min}, y_{i,max}]$. The latter holds because either $\abs{\softplus(m) - x} \geq \delta$, a regime at which $H_\epsilon = H$, or $\abs{\softplus(m) - x} < \delta$, which due to the lipschitzness of $\softplus^{-1}$ it holds that $\abs{m - \softplus^{-1}(x)} \leq L \abs{\softplus(m) - x} \leq L \delta $. Thus, for $\delta < \nicefrac{\epsilon}{2L}$, the claim holds. Similarly, we can use the approximated Heaviside function once more to combine both regimes of $x$, by outputting $H_\delta(x - x_{\mathrm{large}}) x + \left(1 - H_\delta(x - x_{\mathrm{large}})\right) m_T$.

In total, to approximate $\log\abs{z}$ up to $\epsilon$, requires $O\left(\ln^2\left(\frac{m}{\epsilon} \ln\left(\frac{W_{\max}}{f_{\min}}\right)\right)\right)$ nodes, edges, and depth on top of the base NN used to approximate the four non-negative AC.

\subsubsection{Estimating $\arg{z}$}

In this sub-section, we describe the estimation of $\arg{z}$ by softplus networks, building on the approximations of $\ln\abs{z_{\mathrm{re}}}$ and $\ln\abs{z_{\mathrm{im}}}$ described in the previous sub-section.

$\arg{z}$ can be computed according to the following formula:
\begin{align*}
    \arg{z} &= \mathrm{atan2}(z_{\mathrm{im}}, z_{\mathrm{re}}) \\
    &= \begin{cases}
        \arctan\left(\frac{z_{\mathrm{im}}}{z_{\mathrm{re}}}\right) & z_{\mathrm{re}}>0\\
        \arctan\left(\frac{z_{\mathrm{im}}}{z_{\mathrm{re}}}\right) + \pi & z_{\mathrm{re}}<0 \wedge z_{\mathrm{im}}\geq 0\\
        \arctan\left(\frac{z_{\mathrm{im}}}{z_{\mathrm{re}}}\right) - \pi & z_{\mathrm{re}}<0 \wedge z_{\mathrm{im}} < 0\\
        +\frac{\pi}{2} & z_{\mathrm{re}} = 0 \wedge z_{\mathrm{im}} > 0 \\
        -\frac{\pi}{2} & z_{\mathrm{re}} = 0 \wedge z_{\mathrm{im}} < 0 \\
        \textrm{undefined} & z_{\mathrm{re}} = 0 \wedge z_{\mathrm{im}} = 0
    \end{cases}.
\end{align*}
Since we assumed $\abs{z_{\mathrm{im}}},\abs{z_{\mathrm{re}}}>0$, then only the first 3 cases are relevant. Therefore, we can write the formula using the following compact form:
\begin{align*}
    \arg{z} &= \arctan\left(\frac{z_{\mathrm{im}}}{z_{\mathrm{re}}}\right) + H(- z_{\mathrm{re}})\sgn(z_{\mathrm{im}})\pi,
\end{align*}
where $H$ is the Heaviside function. Furthermore, we can rewrite in terms of $\ln\abs{z_{\mathrm{im}}}$ and $\ln\abs{z_{\mathrm{re}}}$:
\begin{align*}
    \arg{z} &= \sgn(z_{\mathrm{re}} z_{\mathrm{im}})\arctan\left(\abs{\frac{z_{\mathrm{im}}}{z_{\mathrm{re}}}}\right) {+} H(\shortminus z_{\mathrm{re}})\sgn(z_{\mathrm{im}})\pi, \\
    &= \sgn(z_{\mathrm{re}} z_{\mathrm{im}})\arctan\exp\left(\ln\abs{z_{\mathrm{im}}} - \ln\abs{z_{\mathrm{re}}}\right) \\
    &\phantom{=} + H(\shortminus z_{\mathrm{re}})\sgn(z_{\mathrm{im}})\pi.
\end{align*}

The signs of $z_{\mathrm{re}}$ can be computed as $\sgn(z_{\mathrm{re}}) = H(o_{\mathrm{re},+} - o_{\mathrm{re},-}) - H(o_{\mathrm{re},-} - o_{\mathrm{re},+})$, which can be approximated with softplus networks using the approximated Heaviside function, $H_\delta$ (defined in previous sub-section). Since we proved in the previous section that $\abs{o_{\mathrm{re},+} - o_{\mathrm{re},-}} \geq \ln \left(\frac{f_{\min}}{\left(m W_{\max}\right)^m} + 1\right)$ then using $0 < \delta < \ln \left(\frac{f_{\min}}{\left(m W_{\max}\right)^m} + 1\right)$ the approximated Heaviside function will be equivalent to the exact Heaviside in the regime of our network. Similarly, $H(-z_{\mathrm{re}}) = \frac{1 - \sgn(z_{\mathrm{re}})}{2}$, and so could be computed exactly as well. The multiplications between these terms and $\arctan \exp (.)$ can be approximated via \textcite{Yarotsky2017}, where in this case the since the values are all bounded by $\pm2$, then we only need $O(\ln(\nicefrac{1}{\epsilon}))$ nodes, edges and depth for this sub-network.

To approximate $t(x) \equiv \arctan \exp (x)$, we start with a piecewise-linear approximation, which can then be approximated to arbitrarily precision with softplus networks. Since $t(x) = \frac{\pi}{2} - t(-x)$, then it is enough to show an approximation for $x \leq 0$, and the $x > 0$ case can be constructed with the above identity. For $x \leq 0$, $t$ is a $\nicefrac{1}{2}$-smooth convex function because its derivative, $\frac{1}{\exp(x) + \exp(-x)}$, is strictly increasing and bounded by $\nicefrac{1}{2}$ in this range. Hence, for any $x,y < 0$ it holds that $\abs{t(x) + t'(x)(y-x) - t(y)} \leq \frac{1}{4} \abs{y - x}^2$.

For any $x_{\min} < 0$ and $n \in \N$, define the following piecewise linear function. Use $n+1$ uniformly spaced anchor points in the $[x_{\min}, 0]$ range, where the first and last anchors are the boundaries. For every anchor point $x$, denote the first-order linear approximation at this point by $l_x(y) = t(x) + t'(x) (y - x)$. Since $t$ is convex in this range, then $l_x(y) \leq t(y)$, and so for every two neighboring anchor points $x_1 < x_2$ the intersection point of $l_{x_1}$ and $l_{x_2}$ must lie in the range $(x_1, x_2)$. Define the segments of the piecewise linear function according to the intersection points and the matching linear approximations $l_x$ of the  anchor point within each segment. For any two neighboring anchors $x_1 < x_2$ and $x_1 \leq y \leq x_2$, this function can be denoted by $\max\{l_{x_1}(y), l_{x_2}(y)\}$. Using the last inequality, we can bound the error in the $[x_{\min},0]$ range with $\frac{1}{4}\left(\frac{x_2 - x_1}{2}\right)^2 \leq \frac{x_{\min}^2}{16 n^2}$. For $x < x_{\min}$, we extend the segment of the anchor $x_{\min}$ until its intersection with the $x$-axis, followed by an open segment for the zero function. Since $\arctan x \leq x$ for any $x > 0$, then $t(x) \leq \exp(x)$, and so for $x_{\min} = -\ln(1/\epsilon)$ it holds that $\forall x \leq x_{\min}, \abs{t(x)} < \epsilon$. Thus, a piecewise linear function with $O\left(\ln(1/\epsilon)\sqrt{\frac{1}{\epsilon}}\right)$ segments can approximate $t(x)$ up to $\epsilon$ maximal difference. Finally, a piecewise linear function with $k$ segments can be realized with a ReLU network of $O(k)$ nodes and edges, and of constant depth.

\section{Proof Sketch for Extending Results to Approximated Contraction Schemes}\label{app:approximated_contractions}

Some TN that cannot be efficiently contracted exactly can still be used with approximate. One of the most notable example for such a case are PEPS. When examining these approximated contraction algorithms, they typically mostly involve iterative application of linear operations, except for employing a Singular-Value Decomposition operation. Hence, if we could simulate the SVD operation with a NN, then we could \emph{trace} the operations of the approximated contraction scheme, use Theorem~\ref{th:main} for simulating the linear operations and then approximate the remaining SVD operations.

To simulate SVD, we can rely on one of the iterative approximation algorithms~\citep{SVDImplementation} used to compute it in practice. This algorithm involves iteratively employing matrix multiplications, computing the $L_2$ norm of a vector, and some divisions, hence the only missing part is approximating divisions and square-root operations (for the $L_2$-norm). Since our construction already represent the log-value of intermediate computations, then performing divisions is just as easy as multiplications~--~simply a subtraction.
As for the square-root, it can be approximated with Newton's method with a quadratic convergence, using just additions, multiplications and divisions. Combining these methods, an SVD can be simulated with NN, and hence some of the most common approximated contraction schemes as well.

\fi

\end{document}